\documentclass[useAMS,usenatbib]{mn2e}
\usepackage{graphicx,amssymb}
\usepackage{times}

\newcommand{\bm}[1]{\mbox{\boldmath $ #1 $}}
\newcommand{\dm}{{\rm d}}
\newcommand{\qm}[1]{``#1''} % Quotation marks
\newcommand{\sm}[1]{`#1'} % Single quotation marks
\title[AGN feedback in cosmological simulations]{The impact of radio feedback from active galactic nuclei in cosmological simulations: Formation of disk galaxies}
\author[T. Okamoto, R. S. Nemmen and R. G. Bower]{Takashi Okamoto$^{1}$\thanks{E-mail:
takashi.okamoto@durham.ac.uk}, Rodrigo S. Nemmen$^{2}$, Richard G. Bower$^1$\\
$^{1}$Institute for Computational Cosmology, Department of Physics, Durham University, South Road, Durham, DH1 3LE\\
$^{2}$Instituto de F\'isica, Universidade Federal do Rio Grande do Sul, Campus do Vale, Porto Alegre, RS, Brazil}
\begin{document}

\date{}

\pagerange{\pageref{firstpage}--\pageref{lastpage}} \pubyear{2007}

\maketitle

\label{firstpage} 
\begin{abstract}
In this paper, we present a new implementation of feedback due to active galactic nuclei (AGN) 
in cosmological simulations of galaxy formation.  
We assume that a fraction of jet energy, which is generated by an AGN, is transferred to 
the surrounding gas as thermal energy. 
Combining a theoretical model of mass accretion onto black holes with 
a multiphase description of star-forming gas, we self-consistently follow 
evolution of both galaxies and their central black holes. 
The novelty in our model is that we consider two distinct accretion modes: 
standard radiatively efficient thin accretion disks and radiatively inefficient 
accretion flows which we will generically refer to as RIAFs; 
motivated by theoretical models
for jet production in accretion disks, we assume that only the RIAF 
is responsible for the AGN feedback.
The focus of this paper is to investigate the interplay between galaxies 
and their central black holes during the formation of a disc galaxy. 
We find that, after an initial episode of bursting star formation, the 
accretion rate onto the central black hole drops so that the accretion
disk switches to a RIAF structure.
At this point, the feedback from the AGN becomes efficient and slightly suppresses star 
formation in the galactic disk and almost completely halts star formation in the bulge. 
This suppression of the star formation regulates mass accretion onto the black hole and 
associated AGN feedback.
As a result, the nucleus becomes a stochastically fuelled low-luminosity AGN (Seyfert galaxy) 
with recurrent short-lived episodes of activity after the star bursts. 
During the \qm{on} events the AGN produces reasonably powerful jets (radio-loud state) 
and is less luminous than the host galaxy, while in the \qm{off} phase the nucleus is 
inactive and \qm{radio-quiet}. 
Our model predicts several properties of the low-luminosity AGN including the bolometric luminosity, 
jet powers, the effect on kpc-scale of the radio jet and the AGN lifetime, 
which are in broad agreement with observations of Seyfert galaxies and their radio activity.
We also find that the ratios between the central black hole mass and the mass of the host 
spheroid at $z = 0$ are $\sim 10^{-3}$ regardless of the strength of either supernova feedback 
or AGN feedback because the radiation drag model directly relates the star formation activity 
in the galactic centre and the mass accretion rate onto the central black hole. 
\end{abstract}


\section{Introduction}

Recent observations have suggested a fundamental connection between 
active galactic nuclei (AGN) and the formation of galaxies. 
Firstly, there is now a well established correlation between the properties
of galaxies and the masses of the black holes (BH) at their centres. The
mass of the central black holes tightly correlates 
with the mass of the host galactic bulges 
\citep[with a median BH-to-bulge mass ratio of $\simeq 0.001$, e.g.][]{kor95, mag98, mer01, mcl02, mar03}, 
as well as with the stellar 
velocity dispersion of the bulges \citep{fm00, geb00, tre02}. 
This discovery points to a fundamental connection between the growth of 
central BHs and the formation of stellar spheroids in galaxies. 
Secondly, high resolution X-ray observations of galaxy clusters have 
revealed large, radio-plasma cavities in intra-cluster medium. These
are usually associated with episodic outbursts from a central radio
galaxy and indicate that huge amounts of mechanical 
energy are being deposited into the intracluster medium by powerful AGN-driven
jets \citep[e.g.][]{all06, bir04, fab06, raf06, tay06}. 
Simulations of the impact of these AGN-driven radio cavities 
suggest that this powerful feedback provides sufficient energy 
  to offset the cooling radiation from the cluster and potentially 
explain why so little cool ($T<3\,{\rm keV}$) 
  gas is seen in these systems  \citep{qui01, chu02, dal04, omm04, sij06, sij07}.  

Motivated by these discoveries, simple prescriptions for the
feedback from AGN (aka. ``radio-mode feedback'') have recently been 
incorporated into semi-analytic galaxy formation models 
\citep[e.g.][]{gra04, mf05, cat05, cro06, del06, bow06}. 
Including radio-mode AGN feedback has resulted in dramatic improvements 
in the models' ability to match the sharp decline
of the galaxy luminosity function and to explain the ``down-sizing''
seen in the evolution of the galaxy population. In particular, 
\citeauthor{bow06} shows that, by assuming that AGN feedback operates only 
in quasi-hydrostatic halos where the cooling time is longer than the 
dynamical time \citep{dal05, sij06},  
the galaxy luminosity functions in local and higher 
redshifts universe can be matched well. \cite{cro06} and \cite{kw07}
showed that similar results were obtained by 
modifying the Bondi-Hoyle-Lyttleton accretion rate formula \citep{hl39, bh44, bon52} 
to account for the multiphase structure of the accreting gas in rapidly cooling halos.

%Nevertheless, while these prescriptions are motivated by theoretical 
%models, it is hugely important to investigate the effect of AGN jets using 
%direct numerical simulations that include a minimum of sub-grid 
%approximation.  Although it is not currently possible to simulate 
%comparably large volumes of the universe in this way, we can see if the 
%successful recipes used in the semi-analytic modelling are justified. 

Recent simulations also have begun to track the impact of AGN feedback
on the galaxy population  \citep{kg05, sij06, dsh05, sdh05a, sij07, dim07}.
One of the first hydrodynamic simulation of galaxy formation which invoked 
AGN feedback was performed by \cite{kg05} by using cosmological simulations 
with smoothed particle hydrodynamics (SPH). 
They reproduced observed X-ray and optical properties of elliptical galaxies 
by injecting thermal energy into the centre of the main progenitor 
at $z < 1$, assuming that sufficient BH mass was present 
for the AGN to be active when a convergent gas inflow exists at the 
centre of the main progenitor. 
A self-consistent model for AGN ``quasar-mode'' feedback associated with 
BH accretion in simulations of galaxy mergers was proposed 
by \cite{dsh05} and \cite{sdh05a}. 
They estimated accretion rate onto BHs by 
using a Bondi-Hoyle-Lyttleton parameterisation 
and injected some fraction of accreted rest mass energy into the 
surrounding interstellar medium (ISM) in the form of thermal energy. 
By using this model and performing a series of merger simulations, 
\cite{hop06} 
simulated evolution of quasar luminosity and predicted luminosity 
functions of quasars.

In this paper, we introduce a new methodology to incorporate BH growth and 
AGN radio feedback self-consistently in cosmological simulations of 
galaxy formation. We are motivated to do this by the recent
semi-analytic models that we have discussed above. These indicate that
the distinction between ``quasar'' and ``radio'' modes is necessary to 
achieve good matches to the global properties of galaxies. 
We distinguish these two modes depending on the accretion rate onto the 
central BHs assuming that the radio-mode exists only when the accretion rate onto the 
BH is low.  

We estimate the mass accretion rate onto the central BHs by combining 
a radiation drag model by \cite{ku02} with our description of the ISM and 
star formation. 
A virtue of the radiation drag model is that it relates star formation activity 
in galactic centre, which can be resolved in our simulations, 
to mass accretion onto a central BHs.  
It implies that a fixed fraction of the star formation in bulges is converted into BH mass 
just as assumed in most semi-analytic models \citep[e.g.][]{cro06, bow06}.  
Since our estimate of the accretion rate onto central BHs is based on radiation 
from stars, the mass growth of BHs in our simulations is resolution independent 
so long as the star formation is modelled in a resolution independent fashion. 

In order to model the AGN radio-mode feedback, we distinguish two 
fundamentally different modes of AGN accretion:
radiatively efficient geometrically thin accretion flows 
\citep[standard disks;][]{ss73} and geometrically thick, radiatively inefficient 
accretion flows \citep[which we will generically refer to as RIAFs;][]
{nmq98, nar05, nem06, yuan07}. We assume only the 
latter are responsible for the radio-mode feedback through production of powerful jets 
\citep[e.g.][]{ree82, mei01, mac03, chu05}. This assertion is
supported by theoretical models for jet production since the jet
power is intimately linked to the scale of the magnetic field loops in the
vicinity of the last stable orbit around the BH 
\citep[e.g.][]{lop99, mei01, nem07}.
Since RIAFs exist only when the accretion rate is much lower than 
the Eddington rate \citep{nmq98}, AGNs can produce strong feedback in objects 
which have low specific star formation rates because the radiation drag model 
directly connects the star formation rate around BHs and mass accretion rate onto the BHs. 
Given the limited numerical resolution inherent in our simulations, we adopt a subgrid model for 
the impact of the AGN feedback. We simply assume that a certain fraction of the available jet power 
thermally couples to nearby diffuse halo gas.

%MAIN POINTS: 
%(a) when disk galaxies are active, they are Seyfert/LINER
%(b) evidence that Sy galaxies host 10-100 pc radio jets
%(c) evidence that Sy galaxies have kpc-scale jets
%(d) these jets may impact the environment of the host galaxy

Until today, most of simulations of galaxy formation with AGN feedback aimed to 
form elliptical galaxies because the AGN feedback is the most plausible 
mechanism that halts the star formation activity in elliptical
galaxies \citep{kg05, sdh05a, dsh05, sij07}. 
Therefore roles of AGN feedback in disk galaxy formation are still largely 
unknown in spite that disk galaxies also harbour central supermassive BHs in 
their bulges. 
In this paper we hence focus on the impact that feedback from an AGN 
jet might have on the formation of a large disk galaxy, and the resulting coevolution 
of the AGN and the host galaxy. 
The majority of AGN hosted by disk galaxies are known to be either Seyfert nuclei or 
the lowest luminosity AGN, the low-ionisation nuclear emission-line regions (LINERs) 
(\citealt{ho97}; see \citealt{ho04} for a review).  
Several targeted observations of individual objects 
(e.g., \citealt{wilson83, morganti98, cecil00, whittle04, keel06, kharb06, middel07}) 
indicate that Seyfert galaxies commonly host compact radio jets spanning $\sim 10-100$ pc 
which strongly affect the state of the ISM and contribute to its ionisation via shocks 
(e.g., \citealt{kukula93,kukula96,falcke98, capetti99,bicknell98,mundell03,riffel06}). 
It is also common that kpc-scale radio structures are observed, which are thought as 
extensions of the inner radio jets 
(e.g., \citealt{wilson83,morganti98,cecil00, whittle04,keel06,kharb06,middel07}), 
and sometimes blow out of the galactic disk with no preferential direction \citep{schmitt01, galli06}.
Very Large Array surveys of samples of Seyfert and LINER galaxies indicate that 
radio outflows with extent $\gtrsim 1$ kpc are a common feature in Seyfert galaxies, 
and presumably are driven by the AGN jets \citep{colbert96,galli06}. 
\citet{galli06} in particular find that $\gtrsim 44 \%$ of the galaxies in their complete sample 
display kpc-scale radio outflows, and they cannot rule out that most Seyfert galaxies produce extended radio outflows.
The kinetic power carried by the Seyfert jets as estimated from observations can be as high as $10^{43} {\rm erg}\ {\rm s}^{-1}$ 
(e.g., \citealt{wilson83,morganti98,kharb06}). 
If an appreciable fraction of this power reaches out of the nuclear region, then the jets may affect the environment 
of the AGN and be a source of feedback in Seyfert galaxies. The impact of feedback from AGN jets on the nuclear 
environment of the host disk galaxies and the evolution of the nuclear activity during the formation of the disk 
galaxy are open problems, which we attack using our cosmological simulations.
In this paper, we show that distinguishing the two modes of AGN accretion yields results which are in broad agreement 
with observations of Seyferts. 

%The formation of disk galaxies is a challenging problem for 
%the cold dark matter model in general and 
%for simulations involving AGN feedback in particular, 
%which provide further reasons for concentrating our efforts 
%on understanding the formation of such systems. One of the difficulties lies
%in conserving the angular momentum of the baryonic content of the 
%dark matter halos as they collapse and form stars. Due to the hierarchical
%growth of the dark matter halo, it is easy to dissipate the angular 
%momentum as satellite galaxies merge into the main progenitor. 
%Furthermore, Okamoto et al.~2005 (hereafter \cite{oka05}) found that any galaxy 
%would become an elliptical galaxy if one injects huge energy into the galactic 
%centre during a burst of star formation. 
%They claimed that their ``density-induced" 
%starbursts were analogous to AGN feedback.  Here we show that
%the role of AGN is considerably more complex, and that distinguishing the
%two modes of AGN accretion yields 
%results which are in broad agreement with observations of Seyferts. 

The paper is organised as follows. We first describe our 
microscopic descriptions of the ISM, star formation, and stellar 
feedback in Section 2 and BH growth and AGN feedback in Section 3. 
We present the details of our cosmological simulations in Section 4 
and the results in Section 5. 
Finally, we summarise and discuss our results in Section 6. 
Readers who are observation-oriented and are not interested in the technical details of 
the simulation may wish to skip directly to Section \ref{obs}, where we compare our results 
with observations of nuclear activity in Seyfert galaxies.

\section{The model of star-forming gas and stellar feedback}

In order to study the effect of AGN feedback on galaxy formation, we  
construct a physically motivated, well-controlled model of the star-forming gas. 
Based on the simplest star formation and feedback recipes used in the 
first generation of hydrodynamic simulations of galaxy formation 
\citep{nw93, sm94, co92, kwh96}, several authors have introduced `multiphase' 
models for star-forming gas in which the ISM is treated as a number 
of distinct phases by formulating differential equations that model 
the interactions between the phases 
(\citealt{yep97}; Springel \& Hernquist 2003 hereafter \citet{sh03}; \citealt{sg03}; \citealt{oka05};
 \citealt{sca06} but see \citealt{bto07} where they explicitly model `clouds' by means of sticky particles 
 in SPH simulations). 
Our model is based on the one used in \cite{oka05} but has 
been significantly modified. 
We describe the details in the following subsections. 

\subsection{Gas cooling}

We compute the non-equilibrium cooling/heating rate and ionisation state 
of each gas particle by using the cooling functions in \cite{the02} which 
include cooling by H, He, and metals, in the presence of an evolving 
but uniform ultraviolet background \citep{hm96} that is switched 
on at $z \simeq 6$.  Inverse Compton cooling is also included. 

Metals are carried by particles and once assigned to a particle, they 
remain with it. 
Nonetheless, effective mixing takes place because we use the 
smoothed metallicity (smoothed in the same way as other SPH quantities) 
when computing the cooling rate and defining the metallicity of a 
newly-formed star. 

\subsection{Cloud formation and the cloud model} \label{SEC:CLOUD}

As in \cite{sh03} and \cite{oka05} we treat an SPH particle as a 
hybrid particle that consists of two distinct phases, i.e. hot 
ambient gas and cold clouds, once its gas density, $\rho$, exceeds a 
density threshold, $\rho_{\rm th}$. 

We assume that the cold clouds form and grow through thermal instability, 
that is
\begin{equation}
\left. \frac{\dm M_{\rm c}}{\dm t} \right|_{\rm TI}
= - \left. \frac{\dm M_{\rm h}}{\dm t} \right|_{\rm TI}
= \frac{\Lambda_{\rm net}(\rho_{\rm h}, u_{\rm h}, Z)}{u_{\rm h} - u_{\rm c}} 
\frac{M_{\rm h}}{\rho_{\rm h}}, 
\label{COOL}
\end{equation}
where $M_{\rm c}$ and $M_{\rm h}$ are mass in cold clouds and mass in the 
hot phase associated with the particle, respectively, $\Lambda_{\rm net}$ 
is the cooling function for gas of metallicity $Z$, and $u_{\rm h}$ 
and $u_{\rm c}$ represent specific internal energy of the hot phase and 
cold clouds, respectively. 
This implies that we assume that the gas is thermally unstable when 
$\rho > \rho_{\rm th}$. 
Cold clouds remain at a fixed temperature, $T_{\rm c} = 100 K$ 
hence $u_{\rm c}$ is a constant as well.
Note, however, as far as $T_{\rm c}$ is much lower than $10^4$ K our 
results are hardly affected by the choice of $T_{\rm c}$. 

We allow cold clouds to have a mass spectrum
\begin{equation}
\phi(m) \equiv \frac{\dm N_{\rm c}}{\dm m} \propto m^{-\alpha}.  
\label{SPEC}
\end{equation}
Observationally $\alpha$ is in a range between $\alpha = 1.5$ to $1.9$ 
\citep{sr89, fuk01, hcs01}.  
Since the effect of changing the value of $\alpha$  is completely degenerate 
with changing other parameters we will introduce such as star formation 
efficiency and thermal conduction efficiency, we simply fix 
$\alpha = 1.7$ and the mass range of clouds, $10^2$--$10^6 M_\odot$, 
throughout this paper.

Following \cite{sg03}, we assume spherical clouds which follow the 
mass-radius relation of \cite{elm89}
\begin{equation}
\frac{m}{r^2} = 190 \left(\frac{M_{\odot}}{\textrm pc^2}\right) P_4^{1/2}, 
\label{MR}
\end{equation}
where $P_4 = \frac{P/k}{10^4 {\rm K}~{\rm cm}^3}$ is the pressure in 
the ambient hot phase which is equal to the effective pressure of
the particle. 

\subsection{Star formation}

We assume that giant molecular clouds in mass range 
of $10^4 M_{\odot}$ to $10^6 M_{\odot}$
are eligible to form stars with a star formation timescale that is 
proportional to the dynamical time of each cloud. 
Therefore the star formation rate for a cloud of mass $m$ is
\begin{equation}
\dot{m}_* = \frac{\dm m_*}{\dm t} = \frac{m}{t_{\rm sf}} = c_* \frac{m}{t_{\rm dyn}}, 
\label{SF}
\end{equation}
where $m_*$ is stellar mass, $t_{\rm sf} = t_{\rm dyn}/c_*$ is the star formation 
timescale for a cloud of mass $m$, $c_*$ is a dimensionless star formation 
efficiency parameter that must be less than unity, 
and $t_{\rm dyn}$ is the dynamical timescale of the cloud. 
For the mass-radius relation in equation~(\ref{MR}), the dynamical timescale of 
a cloud of mass $m$ is given as
\begin{equation}
t_{\rm dyn} = \left[ \frac{3\pi}{32 G \rho(m)}\right]^{1/2}
\simeq 0.32 \  P_4^{-3/8} \left( \frac{m}{M_\odot} \right)^{1/4} {\rm Myr}, 
\end{equation}
where $\rho(m)$ is the mean density of a cloud of mass $m$. 

It should be noted that now the star formation timescale 
is a function of ambient pressure and it could be very 
short in high-pressure medium. 
Therefore we naturally have `shock-induced' starbursts 
that \cite{oka05} introduced by changing the star formation 
efficiency by hand when the rate of change of the entropy by 
shock heating exceeds a threshold value. 
The total star formation rate for clouds of total mass of 
$1 \, M_{\odot}$ with the mass spectrum $\phi(m)$ is obtained by 
integrating $\dot{m}_*$ over the range of masses that are 
eligible to form stars,  
\begin{equation}
\dot{M}_* = \int_{10^4 M_{\odot}}^{10^6 M_{\odot}} \dot{m}_*(m) \phi(m) \dm m, 
\label{TOTSF}
\end{equation}
where the normalisation 
\[\int_{10^2 M_{\odot}}^{10^6 M_{\odot}} m \phi(m) \dm m = 1\]
is applied. 

\subsection{Cloud evaporation}

The different phases of the ISM exchange mass by evaporation. 
As in \cite{sg03}, we use the model of \cite{cm77} for the cloud evaporation, that is,  
\begin{eqnarray}
\dot{m}_{\rm EVP} &=& \left.\frac{\dm m}{\dm t}\right|_{\rm EVP} = \eta_{\rm EVP} 
\frac{16 \pi \mu \kappa r}{25 k_{\rm B}} \nonumber \\
&\simeq& 4.4 \times 10^{-16} \eta_{\rm EVP} \left(\frac{r}{\rm pc}\right) 
T^{5/2} \ \left( \frac{m}{M_{\odot}} \right) M_{\odot} \ {\rm Myr}^{-1} \nonumber \\ 
&\propto& m P_4^{-1/4} T^{5/2}, 
\label{EVP}
\end{eqnarray}
where $\eta_{\rm EVP}$, $\mu$, $\kappa$, $r$, and $k_{\rm B}$ 
are the dimensionless conduction efficiency parameter, 
the mean molecular weight, the conductivity where we use the Spitzer 
value \citep{spi62}, the radius of the cloud, and the Boltzmann constant.  
The total evaporation rate for clouds of total mass $1 \, M_{\odot}$ then becomes 
\begin{equation}
\dot{M}_{\rm EVP} = \int_{10^2 M_{\odot}}^{10^6 M_\odot} \dot{m}_{\rm EVP}(m) \phi(m) \dm m. 
\label{TOTEVP}
\end{equation}

\subsection{Supernova feedback and chemical enrichment}

We consider each stellar particle as a single stellar population (SSP) 
having its own age, metallicity and initial mass function (IMF) $\phi_*(m)$. 
Throughout this paper we assume that the IMF is always the Salpeter IMF 
\citep{sal55} whose lower and upper mass limits are 
$0.1$ and $100$ $M_{\odot}$, respectively. 
The recycled mass fraction, type II and type Ia supernova (SN) rates, 
and metal production rates for some species are calculated according to 
the age, metallicity, and IMF of the SSP. 
Each SN is assumed to give $\eta_{\rm SN} 10^{51}$ erg of energy to surrounding 
gas, where $\eta_{\rm SN}$ is the feedback efficiency parameter. 
We refer the readers to \cite{oka05} and references therein for a more detailed 
description. 

\subsection{Numerical implementation}

We have implemented the physics described above into a publicly available 
PM-TreeSPH code GADGET2 \citep{gadget2}, a successor of the TreeSPH 
code GADGET \citep{gadget1}. 
We do not use the instantaneous recycling approximation (IRA) which is 
often assumed in studies of galaxy formation. 
In this section we describe how the physical processes described above are 
implemented in our code. 

For $\rho > \rho_{\rm th}$, an SPH particle is eligible to form cold 
clouds. Hence it is eligible to form stars.  
We compute the probability $p_*$ of a SPH particle spawning 
a new star particle of mass $m_*$ during a time-step $\Delta t$ as  
\begin{equation}
p_* = \frac{M_{\rm c}}{m_*} \left[1 - \exp\left(-\frac{\Delta t}{t_*}\right)\right],
\end{equation}
where the average star formation timescale $t_*$ is obtained by
using equation (\ref{TOTSF}):
\begin{equation}
t_*(P) = \left(\frac{\dot{M}_*(P)}{M_{\odot}}\right)^{-1}  {\rm Myr}. 
\end{equation}
We use $m_* = m_{\rm orig}/3$ as in \cite{oka05}, where 
$m_{\rm orig}$ denotes the original SPH particle mass. 
When an SPH particle spawns a star particle, the mass in cold clouds is 
reduced by $m_*$. 

We relax the IRA through a probabilistic treatment of SNe II and Ia. 
First, we assume that stars more massive than $8 M_{\odot}$ explode 
as SNe II. 
The probability $p_{\rm II}$ of a star particle having an event of SN II 
explosion during a time-step $\Delta t$ is given by 
\begin{equation}
p_{\rm II} = \frac{\int_{t}^{t + \Delta t} r_{\rm II}(t') \dm t'}
{\int_{t}^{t(8 M_{\odot})} r_{\rm II}(t') \dm t'}, 
\end{equation}
where $t$ is the age of the SSP, $r_{\rm II}(t)$ is the SN II rate for the SSP of age $t$ 
and $t(8M_{\odot})$ is the lifetime of the star of original mass $8 M_{\odot}$.  
When a star particle satisfies the condition to have an event of SN II we distribute 
mass, metals, and feedback energy, which are expelled by 
the total SNe II between $t = 0$ and $t(8 M_{\odot})$ to the surrounding gas particles 
using the SPH kernel weighting.  

Since the SN Ia rate is much lower than the SN II rate, we allow star particles to have only 
one SN Ia explosion event for each SN Ia time-step $\Delta t_{\rm Ia}$, 
which is much longer than the time-steps for dynamical calculation. 
By doing this, we substantially reduce the 
computational expense that is needed mainly for the neighbour search to distribute 
energy, mass, and metals.   
As for SN II, we can compute the SN Ia probability during a time-step $\Delta t$ as
\begin{equation}
p_{\rm Ia} = \frac{\int_t^{t + \Delta t} r_{\rm Ia}(t') \dm t'}
{\int_{t}^{t_0 + \Delta t_{\rm Ia}} r_{\rm Ia}(t') \dm t'}, 
\end{equation}
where $t_0$ is the age at which the previous SN Ia time-step finished and 
$r_{\rm Ia}(t)$ is the SN Ia rate for a SSP of age $t$. 
We adopt $\Delta t_{\rm Ia} = 100$ Myr in this paper. 
The probabilistic method described above statistically relaxes the IRA. 

Finally, we update the mass of the hot phase $M_{\rm h}$ and the specific
internal energy of the hot phase $u_{\rm h}$.  
The new mass of the hot phase ${M_{\rm h}}'$ is given by solving the thermal 
energy equation implicitly using equation (\ref{COOL}), 
\begin{equation}
{M_{\rm h}}' = M_{\rm h} + \Delta M_{\rm EVP} 
- \frac{\Lambda_{\rm net}({\rho_{\rm h}}', u_{\rm h}, Z)}{u_{\rm h} - u_{\rm c}}
\frac{{M_{\rm h}'}}{{\rho_{\rm h}}'} \Delta t, 
\end{equation}
where $\Delta M_{\rm EVP}$ is the mass which evaporates from cold clouds 
during the time-step, and the new density of the hot phase ${\rho_{\rm h}}'$ 
is given by the mass conservation, $M_{\rm SPH} = M_{\rm h} + M_{\rm c}$, 
and by the assumption that the volume occupied by a SPH particle remains the same 
during the time-step, i.e.
\[\frac{M_{\rm SPH}}{\rho} = {\it Vol}_{\rm c} + \frac{M_{\rm h}}{\rho_{\rm h}}, \] 
where ${\it Vol}_{\rm c}$ is the volume occupied by the cold clouds computed 
using the mass-radius relation (eq.~\ref{MR}). 

In addition to the adiabatic heating/cooling and shock heating, non-hydrodynamic 
processes such as SN feedback and cloud evaporation also change the 
specific energy of the hot phase. 
The new specific energy of the hot phase ${u_{\rm h}}'$ due to 
the non-hydrodynamic processes is also calculated implicitly as
\begin{equation}
{u_{\rm h}}' = u_{\rm h} 
+ \frac{\Delta Q + (u_{\rm c} - {u_{\rm h}}') \Delta M_{\rm EVP}}{{M_{\rm h}}'}, 
\label{UHOT}
\end{equation}
where $\Delta Q$ is the thermal energy received by the SPH particle during the 
time-step. 

Note however that we model the SN feedback not as the thermal 
heating of the ISM but as the kinetic heating described in the 
following subsection as {\it winds}. 
This is motivated by the results from high-resolution 2D and 3D simulations of 
the ISM \citep{wn01, wad01, wn07} in which they showed that the ISM is supported 
by the turbulent motion originated in the self-gravity of the gas and galactic 
rotation. They found that energy injection by SNe hardly changes the structure of 
the ISM. 
\cite{kra03} also indicated that the energy feedback from SNe does not change 
the structure of the ISM but significantly reduces the amount of gas in the
galaxy in their cosmological simulation performed with an adaptive mesh refinement 
code. 

Lacking the heating by the $\Delta Q$ term, the ISM loses its pressure quickly 
and would be fragmented. 
We therefore introduce {\it minimal heating} 
by imposing the minimum pressure for the star forming gas 
as a function of density, 
\begin{equation}
P_{\rm min}(\rho) \propto \rho^{\gamma_{\rm eff}} \ {\rm for} \ \rho > \rho_{\rm th}, 
\label{PEFF}
\end{equation}
by which we mimic the heating from self-gravity and galactic shear motion claimed 
by \cite{wn01}, \cite{wad01}, and \cite{wn07}. 
We adopt in this paper the value $\gamma_{\rm eff} = 1.4$ for the effective adiabatic index, 
which is close to the value 
found by \cite{wn07} for dense gas and is stable against gravitational instability
(the critical value is 4/3). 
We normalise the minimum pressure as 
$P_{\rm min}(\rho_{\rm th}) = (\gamma -1) u_4 \rho_{\rm th}$, where $u_4$ denotes 
the specific internal energy of the gas with temperature $10^4$ K. 
Once pressure from the hot phase drops below $P_{\rm min}$, we set the temperature 
of the hot phase to $T_{\rm hot}$ and recalculate the mass in cold clouds so that 
the ambient pressure becomes $P_{\rm min}$. 
We set $T_{\rm hot}$ to $10^6$ K in this paper but our results are hardly affected 
by the value of $T_{\rm hot}$ as far as $T_{\rm hot} >> 10^4$ K. 
In quiescent disks, most of the star forming gas has $P_{\rm min}$ by construction. 
However, for example during a violent merger, the pressure can be significantly higher 
than $P_{\rm min}$ due to shock heating and hence the gas can have very short star formation 
timescale. 

\subsection{Winds}

As discussed in \cite{sh03}, we explicitly assume that 
cold clouds and hot ambient medium remain tightly coupled in high-density 
regions. Hence, there is no obvious route for the high entropy gas to escape from 
the star forming regions. 
We thus extend our feedback model to account for galactic winds by SN feedback. 
Our model is based on \cite{sh03} and is modified to take the non-instantaneous 
SN explosions into account. 

First, we specify the velocity of the wind, $v_{\rm w}$, when it leaves the disk. 
Then, we assign the probability $p_{\rm w}$ with which the gas particle is added 
to the wind as
\begin{equation}
p_{\rm w} = \frac{\frac{1}{2} M_{\rm SPH} {v_{\rm w}}^2}{\Delta Q}, 
\end{equation}
where $M_{\rm SPH}$ is the mass of the gas particle and $\Delta Q$ is 
the feedback energy received during the time-step. 
By doing this, our model becomes insensitive to the numerical 
resolution, time-step in the simulations, and the number of neighbours 
we use to distribute the feedback energy. 
When a particle is added to the wind, we modify the velocity ${\mathbf v}$ of the particle 
according to 
\begin{equation}
{\mathbf v}' = {\mathbf v} + v_{\rm w} {\mathbf n}. 
\end{equation}
For the unit vector ${\mathbf n}$, we choose random orientation along the direction 
${\mathbf v} \times {\mathbf a}_{\rm grav}$, where ${\mathbf a}_{\rm grav}$ 
is the gravitational acceleration.
Therefore wind particles are preferentially ejected along the rotation 
axis of a spinning object as the `axial wind' in \cite{sh03}.  
We also decouple a new wind particle from hydrodynamic interactions 
for a brief time such that the winds originate from a region close 
to the surface of the star forming region. 
The full hydrodynamic interactions are enabled again once the 
density of the particle has fallen to 0.1 $\rho_{\rm th}$, 
or once a time of $\Delta t$ = 20 Myr has elapsed, 
whichever happens earlier.

\subsection{Parameter setting and test simulations}

In this paper we use the following parameters for star formation 
and SN feedback as a fiducial model; the threshold density, 
$\rho_{\rm th} = 5\times10^{-25}$ g cm$^{-3}$, the dimensionless 
star formation efficiency, $c_* = 0.005$, the dimensionless 
conductivity efficiency, $\eta_{\rm EVP} = 0.1$, the SN feedback 
efficiency, $\eta_{\rm SN} = 1$, and the wind velocity 
$v_{\rm w} = 500$ km s$^{-1}$. 
The relatively low value of $c_*$ compared to other papers comes 
from the fact that we use the dynamical time of each cloud not 
the dynamical time for SPH particles. 
We choose the values of $\rho_{\rm th}$ and $c_*$ so that our 
model reproduce the observed relation between surface gas density 
and surface star formation rate density \citep{ken98}. 
The most important parameters in our model are the feedback 
efficiency and the wind velocity. 
The wind velocity we use is comparable to the strong wind model 
in \cite{nsh04}. 
For our choice of the parameters and the IMF, the wind production 
rate by SNe II is $\dot{M}_{\rm w}/\dot{M}_*  \simeq 3 $. 
There will be further contribution from SNe Ia at later times. 

In this subsection, we show the behaviour of our model in idealised 
simulations of disk formation from virialised gas in the external 
NFW potential \citep{nfw96}. 
Here, $M_{\rm vir}$ is the virial mass of the system and we always 
assume that 10 per cent of the mass is in baryonic form. 
The initial angular momentum in terms of the spin parameter,
$\lambda = J|E|^{1/2}/(GM_{\rm vir}^{2/5})$, is set to 0.07 with 
the assumption that the specific angular momenta $j(r)$ of spherical 
shells are all aligned and their magnitude is given by 
\[j(r) \propto \frac{M(r)}{M_{\rm vir}}.\] 

\begin{figure} 
\includegraphics[width=7.5cm]{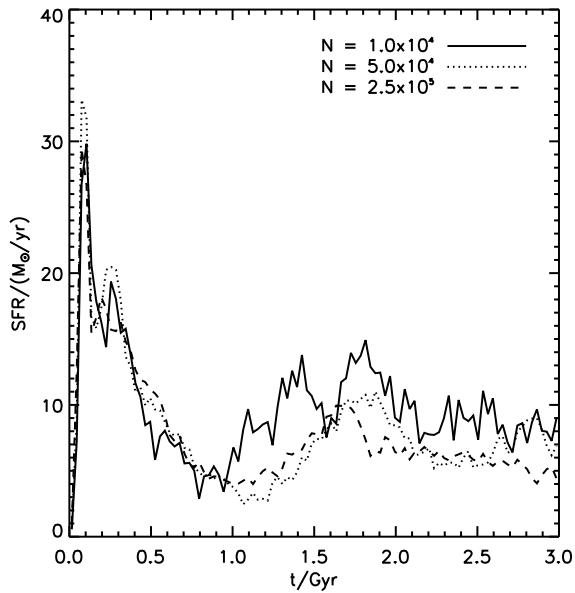} 
\caption{Star formation rates in simulations of disks growing within halos of 
$M_{\rm vir} = 10^{12} M_{\odot}$, with different resolutions.  
The simulations with $N_{\rm gas} =$ 10000, 50000, and 250000 are 
indicated by the solid, dotted, and dashed lines, respectively.
}  
\label{RESOLUTION} 
\end{figure}

In Fig.~\ref{RESOLUTION} we show the star formation rates in the 
simulations for  $M_{\rm vir} = 10^{12} M_{\odot}$ with 
three different resolutions: $N_{\rm gas} =$ 10000, 50000, and 
250000.
The star formation rates quickly converge. 
The simulation with $N_{\rm gas} = 50000$, which is a comparable resolution that 
we will use in cosmological simulations, has very similar star formation history 
to that in the simulation with $N_{\rm gas} = 250000$.
Hereafter we thus only show the simulations with $N_{\rm gas}$ = 50000.

Now we show how star formation rates depend on the values of 
feedback parameters, $\eta_{\rm SN}$ and $v_{\rm w}$. 
We here run the simulations with two additional models:
the slow wind model ($\eta_{\rm SN} = 1$ and 
$v_{\rm w} = 250$ km s$^{-1}$) and the weak feedback model  
($\eta_{\rm SN} = 0.5$ and $v_{\rm w} = 250$ km s$^{-1}$). 
The wind speed in the slow wind model is half of that in the fiducial 
model, therefore it has an uncomfortably high mass loading factor 
$\eta_{\rm w} \simeq 12$. 
Of course the mass loading in the weak feedback model becomes half. 

\begin{figure} 
\includegraphics[width=7.5cm]{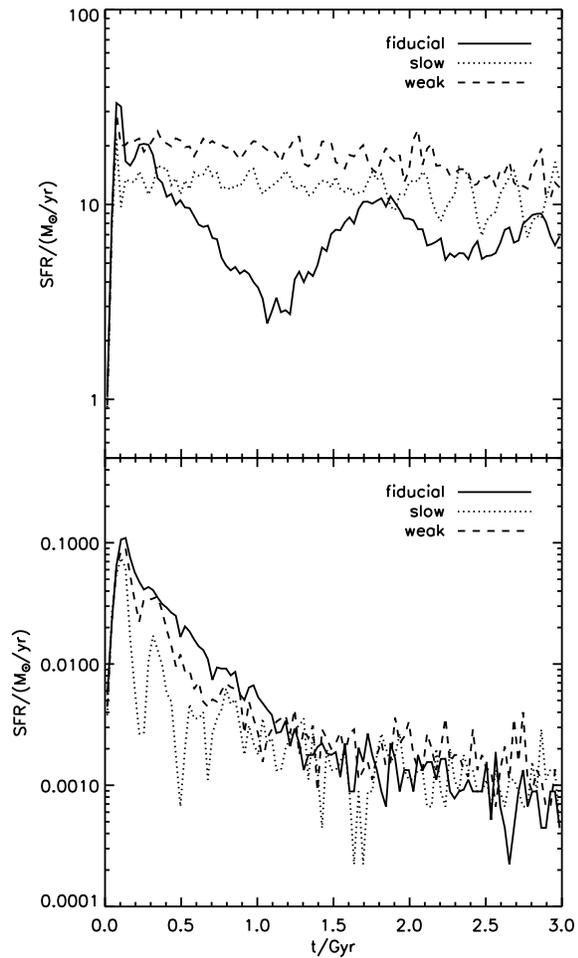} 
\caption{Star formation rates in simulations of disks growing within halos of 
$M_{\rm vir} = 10^{12} M_{\odot}$ (upper panel) and 
$M_{\rm vir} = 10^{10} M_{\odot}$ (lower panel). 
The solid, dotted, and dashed lines indicate the fiducial, slow wind, and 
weak feedback models, respectively. 
}  
\label{SFRWIND} 
\end{figure} 
Fig.~\ref{SFRWIND} shows the star formation rates in galaxies with these 
three wind models for halos 
of $M_{\rm vir} = 10^{12} M_{\odot}$ and $10^{10} M_{\odot}$.
Since $v_{\rm w} = 500$ km s$^{-1}$ exceeds the escape velocity of the 
halo with  $M_{\rm vir} = 10^{12} \ M_{\odot}$ and 
$v_{\rm w} = 250$ km s$^{-1}$ does not, 
  the star formation is more strongly suppressed in the fiducial 
model in spite of the same feedback efficiency.  
We find that the star formation rates of the slow wind and weak feedback models  
have regularly time spaced peaks of activity.  
For the adopted NFW potential, the wind particles with the initial velocity, 
  $v_{\rm w} = 250$ km s$^{-1}$, will  return to the disk in $\sim 0.13$ Gyr.  
  This time-scale might be responsible for the periodicity seen in the star formation rates.  
The star formation rate is highest in the weak feedback model for this halo as expected.  
However situation changes for the halo of 
$M_{\rm vir} = 10^{10} \ M_{\odot}$.
Now the mass loading factor is more important than the wind velocity because 
both wind velocities exceeds the escape velocity of this halo. 
Hence the star formation rate is highest in the fiducial model and lowest 
in the slow wind model although all models have star formation rates 
$\sim 10^{-3} \; M_\odot \, {\rm yr}^{-1}$ after 1 Gyr. 
Since we have to suppress star formation even in relatively large halos 
to form disk galaxies \citep{oka05}, we employ the fiducial parameter set 
in our cosmological simulations.

\begin{figure} 
\includegraphics[width=7.5cm]{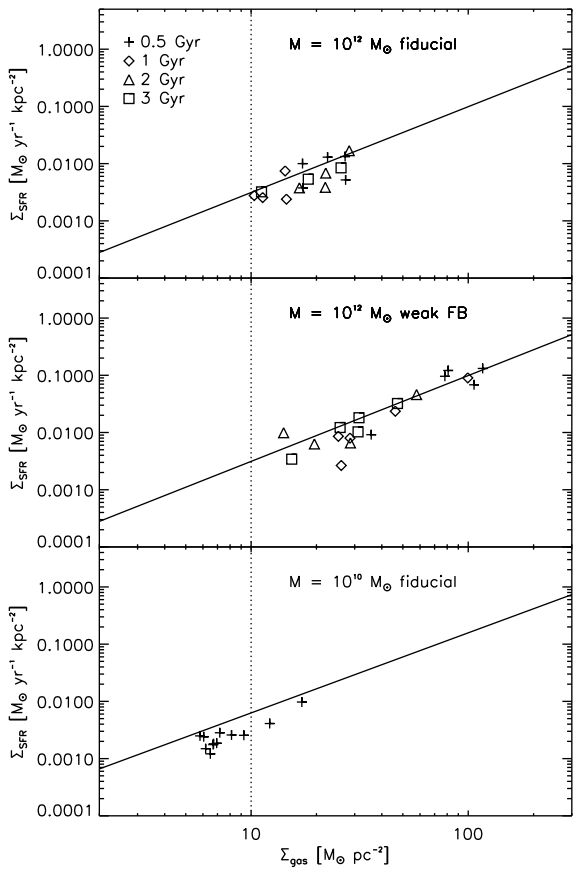} 
\caption{Star formation rate per unit area versus gas surface density
in an idealised simulation of disk formation in a halo with 
$M_{\rm vir} = 10^{12} M_{\odot}$ (top and middle panels).  
The surface star formation rate densities are computed using the
surface density of stars younger than $3 \times 10^{7}$ yr in
cylindrical bins. 
The symbols represent star formation rates at
$t= 0.5, \ 1, \ 2, \ 3$ Gyr, indicated by plus signs, diamonds,
triangles, and squares respectively. 
The top and middle panels respectively show the fiducial and weak feedback models. 
The galaxy in the halo with $M_{\rm vir} = 10^{10} M_\odot$ is shown in the 
bottom panel, and there both the surface star formation rate density and 
the surface gas density are averaged over the galaxy. Points represent the 
evolution of the relation from 0.25 Gyr to 3 Gyr with a time-step of 0.25 Gyr. 
The solid and dotted lines indicate the target relation and 
the cut-off surface density
respectively.
}  
\label{KENNICUTT} 
\end{figure}

In Fig. \ref{KENNICUTT} we show the relations between the star formation rate  
per unit area and the surface gas density for the fiducial (top panel) and 
weak feedback (middle panel) models.  
We estimate the surface star formation rate densities by computing the
surface density of stars younger than $3 \times 10^{7}$ yr in
cylindrical bins. 
Both models show reasonably good agreement with the target relation
\citep[][; the target relation we show is eq.~(25) of \citealt{sh03}]{ken98}.
The use of different wind parameters do not affect this relation in our simulations,  
while the ISM can have higher gas density in the weak feedback model. 
For the galaxy in $M_{\rm vir} = 10^{10} M_\odot$ halo, winds are so effective that 
there are only a handful of young stars in the galaxy. 
We hence only show, in the bottom panel of Fig.~\ref{KENNICUTT}, 
the relation between the global surface gas density and the global surface star formation rate 
, which are averaged over the galaxy.  
The different points in this panel correspond to the relations at different epochs. 
We find that the global relation in the $10^{10} M_{\odot}$ halo is also in reasonable agreement 
with the target relation. 

\section{Black hole accretion and AGN feedback}

\subsection{Seed black holes}

In the cosmological simulations we run a friends-of-friends halo-finder 
\citep{dav85} 200 times on the fly from $z = 15$ to $0$.
When we find a halo that contains more than $N_{\rm th}$ dark matter 
particles and no BH particles, we put a seed BH on a density 
peak of the stellar distribution in the halo. 
We assume 260 $M_{\odot}$ as the mass of the seed black hole. 
This mass is consistent with the theory of stellar evolution, which shows 
that the nuclear burning in very massive stars above 
260 $M_{\odot}$ is unable to halt gravitational collapse
\citep[e.g.][]{heg03}. 
Note that as far as the seed BH mass is sufficiently small, 
our results hardly change because we do not impose an upper limit 
on the mass accretion rate. 
Throughout this paper we use $N_{\rm th} = 128$. 
The simulation employing $N_{\rm th} = 256$ yields almost the 
same results except for the total number of BHs. 
The insensitivity on the choice of these parameters relies on our AGN 
feedback model that is only effective when the mass accretion rate onto the BH
is much lower than the Eddington rate (see \textsection \ref{sec:jetpower}). 

\subsection{Mass accretion by the radiation drag}

One mechanism yielding the proportionality of the BH mass to its host 
bulge mass has been proposed by \cite{ume01} and \cite{ku02}. 
According to \citeauthor{ku02}, in the central regions of galaxies 
the drag due to stellar radiation may result in a loss of angular 
momentum of a clumpy ISM and gives rise to mass inflow toward the 
centre. 
The optical depth of a gas cloud is 
$\bar{\tau} = \chi_{\rm d} \rho_{\rm c} r_{\rm c} \simeq \chi_{\rm d} m_{\rm c}/r_{\rm c}^2$, 
where $\rho_{\rm c}$, $m_{\rm c}$, and $r_{\rm c}$ are the density, mass, and a size of 
a cloud. 
In principle we can use the cloud model we described in \S\ref{SEC:CLOUD}. 
We however assume that the clouds are all identical and randomly distributed in 
a star forming region as in \citeauthor{ku02} for simplicity.  
The mass extinction coefficient $\chi_{\rm d}$ is given by 
$\chi_{\rm d} = 300 \ {\rm cm}^2 {\rm g}^{-1} (a_{\rm d}/0.1{\rm \mu m}^{-1}) 
(\rho_{\rm s}/{\rm g} \ {\rm cm}^{-3})(Z/0.3 Z_{\odot})$, 
where $a_{\rm d}$ is the grain radius, $\rho_{\rm s}$ is the density of solid 
material within the grain \cite[e.g.][]{spi78}, and $Z$ is the metallicity of gas. 
We use $a_{\rm d} = 0.1 \ \mu {\rm m}$, $\rho_{\rm s} = 1$ g cm$^{-3}$, and $Z$ 
is directly obtained in the simulations as the mean metallicity of the gas in a 
star forming region. 
The total optical depth of a star forming region can then be approximated as 
\begin{equation}
\tau_{\rm SFR}(t) \simeq \frac{3 \chi_{\rm d}}{4 \pi} \frac{M_{\rm c}(t)}{R_{\rm SFR}(t)^2}, 
  \label{TAU}
\end{equation}
where $R_{\rm SFR}$ and $M_{\rm c}$ are the size of a star forming region 
and the total mass of the clouds in the star forming region. 

The mass accretion rate due to the radiation drag mechanism described above is 
well approximated by 
\begin{equation}
\dot{M}_{\rm drag} = \eta_{\rm drag} \frac{L_{\rm SFR}(t)}{c^2}(1 - e^{-\tau_{\rm SFR}(t)}), 
  \label{DRAG}
\end{equation}
where $L_{\rm SFR}$ is the total luminosity of stars in the star forming region. 
We calculate $L_{\rm SFR}$ by summing up the bolometric luminosities of all 
stars, each of which is obtained as a function of its age from a look-up table 
generated by P$\acute{\rm E}$GASE2 \citep{pegase}. We here ignore the metallicity 
dependence of the stellar luminosity and use a table for the solar metallicity 
because the dependence on metallicity is weak. 
\citeauthor{ku02} found that the efficiency $\eta_{\rm drag}$ is maximally 
$0.34$ in the optically thick regime assuming a spherical star forming region 
with a uniform density. 
In principle $\eta_{\rm drag}$ can be much larger or smaller depending on the 
geometry and density structure of the star forming region 
(Kawakatu private communication).
We thus change the efficiency and adopt $\eta_{\rm drag} = 1$ 
in this paper, considering more centrally concentrated density distribution 
in real star forming regions than assumed in \citeauthor{ku02}.

Once we define the radius of a star forming region around a BH, $R_{\rm SFR}$, 
we obtain $\chi_{\rm d}$, $M_{\rm c}$, and $L_{\rm SFR}$ 
in equations~(\ref{TAU}) and (\ref{DRAG}) 
directly in simulations, and hence we can compute 
$\tau_{\rm SFR}$ and $\dot{M}_{\rm drag}$ self-consistently; the effects of 
outflows are also implicitly incorporated in the estimate of these variables because 
we calculate them self-consistently at each time-step of the simulation. 
In order to define the size of the star forming region, $R_{\rm SFR}$ 
we start from a sphere around a BH that contains 40 SPH particles, and then 
we search the radius of the sphere that maximises $\dot{M}_{\rm drag}$  
by increasing the radius of the sphere. We employ this radius as the size 
of the star forming region. 
When the mean gas density of the initial sphere is lower than the 
threshold density for star formation, $\rho_{\rm th}$, 
we consider that there is not enough gas around the BH to fuel accretion 
and we simply set $\dot{M}_{\rm drag} = 0$. 
\cite{ku04} argued that radiation from bulge stars contributes to the 
mass accretion to the central hole more efficiently than that from 
disc stars. We crudely mimic this effect by using spherically averaged values 
to define star forming region; by doing this we put less weight to the star 
forming gas that has a flattened distribution. We will show how star formation
in the disc and bulge correlate with the mass accretion rate, $\dot{M}_{\rm drag}$, 
in our simulations later.

Strictly speaking, one should distinguish $\dot{M}_{\rm drag}$ from the accretion rate 
onto the BH, $\dot{M}$, because the radiation drag is not likely to remove the 
angular momentum thoroughly; some residual angular momentum will terminate the radial 
contraction of the accreted gas \citep{sat04}. 
\cite{gra04} introduced the viscous timescale in which the mass accumulated at 
the galactic centre by the radiation drag is fed to the BH. 
We however assume $\dot{M} = \dot{M}_{\rm drag}$ for simplicity. 
The time delay in mass fuelling introduced by the viscous timescale  
may become important when we compare the luminosity of AGNs and their 
host galaxies in detail. 

\subsection{AGN feedback} \label{sec:jetpower}

Motivated by the success of recent semi-analytic models 
\citep{cro06, bow06}, we assume that AGNs directly heat diffuse hot 
gas confined in dark halos through the production of jets. 
As the jet production mechanism, we employ the model proposed by 
\cite{mei01} which is an hybrid version of 
the Blandford-Znajek \citep{bz77} and the Blandford-Payne \citep{bp82} processes. 
In this hybrid model the jet power is generated by the magnetic field threading the 
accretion flow inside and outside the the ergosphere of the BH. 
This model is able to draw upon the rotational energy of the accretion flow as well 
as the spinning black hole in order to drive jets, and its basic features are supported 
by recent relativistic numerical simulations of jet formation \citep[e.g.][] {haw06}.
We refer the readers to \cite{mei01} and \cite{nem07} for more details.

\cite{mei01} presented models both for non-spinning (Schwarzschild) 
and for spinning (Kerr) BHs. 
Semi-analytic cosmological simulations of the spin evolution of BHs 
through mergers and gas accretion \citep{vol05} and estimates of 
the radiative efficiencies of global populations of quasars based on 
Soltan-type arguments \citep[e.g.][]{sol82, yt02, wan06} suggest 
that most nearby massive BHs are rapidly rotating. 
Recently \cite{nem07} also reported that 
BHs in nearby elliptical galaxies are likely to have high spins $j \simeq 0.7-1$.
We thus only consider spinning BHs in this paper. 
\cite{mei01} parameterised the jet luminosity for standard thin
discs as 
\begin{eqnarray} \label{eq:ljet_sda}
L_{\rm jet}^{\rm SD} &\approx &10^{42.7} {\rm erg}\ {\rm s}^{-1} 
\left(\frac{\alpha_{\rm SD}}{0.01} \right)^{-0.1} m_9^{0.9}
\left(\frac{\dot{m}}{0.1}\right)^{6/5} \nonumber \\
    &\times& (1 + 1.1 j + 0.29 j^2), 
  \label{SD}
\end{eqnarray}
and for RIAFs as 
\begin{eqnarray} \label{eq:ljet_riafa}
L_{\rm jet}^{\rm RIAF} &\approx& 10^{45.1} {\rm erg}\ {\rm s}^{-1}
\left(\frac{\alpha_{\rm RIAF}}{0.3}\right) m_9 
\left(\frac{\dot{m}}{0.1}\right) g^2 \nonumber \\
&\times& (0.55 f^2 + 1.5 f j + j^2), 
  \label{RIAF}
\end{eqnarray}
where $\alpha_{\rm SD}$ and $\alpha_{\rm RIAF}$ are 
the viscosity parameters for standard thin discs and for 
RIAFs, respectively, $m_9$ is the black hole mass in units of 
$10^9 M_{\odot}$, 
$\dot{m} \equiv \dot{M}/\dot{M}_{\rm Edd}$ 
is the accretion rate scaled to the Eddington rate assuming the 10\% 
energy conversion efficiency ($\dot{M}_{\rm Edd} \equiv 22 \ m_9 \ M_\odot \ {\rm yr}^{-1}$), 
and $f$ and $g$ are the ratios of the 
actual angular velocity and the azimuthal magnetic field to those 
calculated by \cite{ny95} respectively.  
According to \cite{mei01} we employ $f = 1$ and $g = 2.3$. 
Since the limited numerical resolution of our cosmological simulations 
  prevents us from resolving the details of the propagation of jets and its 
    impact on the surrounding medium, we simply assume that a fraction of the 
    jet energy $\Delta E_{\rm FB}^{\rm jet}  = \eta_{\rm AGN} L_{\rm jet} \Delta t$ 
    is delivered as thermal energy to the neighbouring 40 diffuse gas particles 
that have density, $\rho < 0.1 \rho_{\rm th}$. 
In our simulations, the size of the region heated by the AGN feedback is typically $\sim 10$ kpc.
Unless otherwise stated, 
we use $\alpha_{\rm SD} = \alpha_{\rm RIAF} = 0.1$, $j = 0.5$, 
and $\eta_{\rm AGN} = 0.1$. The adopted value of $j$ corresponds to the intermediate case 
between Schwarzschild (no-spin) and maximally rotating BHs.

As previously mentioned, we adopt two distinct regimes of accretion flows: 
standard thin discs (optically thick, geometrically thin, radiatively efficient) and 
RIAFs (optically thin, geometrically thick, radiatively inefficient). 
The parameter which controls the state of the accretion flow is $\dot{m}$, 
and the critical value of the accretion rate which sets the division between 
the two regimes is given by $\dot{m}_{\rm crit} \approx \alpha^2$ 
\citep[e.g.][]{nmq98}. 
Since the RIAF solution ceases to be valid above 
$\dot{m}_{\rm crit}$, for $\dot{m} \le \dot{m}_{\rm crit}$ a RIAF exists 
and for $\dot{m} > \dot{m}_{\rm crit}$ a thin disc occurs. 
As a consequence of the adopted theoretical model of jet production, 
we have strong AGN feedback only for RIAFs and the maximum AGN feedback occurs 
at $\dot{m} = \dot{m}_{\rm crit}$. 
For the parameters we chose, we have the following jet production efficiency in RIAFs, 
\begin{equation} \label{eq:ljet_riafb}
L_{\rm jet}^{\rm RIAF} \approx 2.6 \times 10^{-1} \dot{M} c^2 \ {\rm for} \ \dot{m} \le \dot{m}_{\rm crit},  
\end{equation}
which is much higher than that in standard discs, given by  
\begin{equation} \label{eq:ljet_sdb}
L_{\rm jet}^{\rm SD} \approx  8.1 \times 10^{-5} \dot{M} c^2 \ {\rm for} \ \dot{m} > \dot{m}_{\rm crit},   
\end{equation}
where in equation~(\ref{eq:ljet_sdb}) we adopt $m_9 = \dot{m} = 1$. The equations (\ref{eq:ljet_riafb}) 
and (\ref{eq:ljet_riafa}) correspond to the 
\qm{radio-loud mode} in our simulations, and the equations (\ref{eq:ljet_sdb}) and (\ref{eq:ljet_sda}) 
represent the \qm{radio-quiet mode}.

Note that, in principle, we could have even higher efficiencies of jet production, because the 
energy is drawn from the \qm{spin energy} stored in the BH. We explore the case of 
a higher efficiency corresponding to a higher BH spin later in \textsection \ref{PARAMETERS}.

\section{Cosmological simulations of disc galaxy formation with AGN feedback}

To study the role of AGN feedback in disk galaxy formation, 
we use the initial conditions presented in \cite{oka05}, from which 
they produced an extended disk galaxy.
The background cosmology is the so-called $\Lambda$CDM with 
the mean matter density $\Omega_0 = 0.3$, Hubble parameter, 
$h \equiv H_0 /100$ km s$^{-1}$Mpc$^{-1}=0.7$, cosmological constant 
term, $\Omega_\Lambda \equiv \Lambda_0/(3 H_0^2) = 0.7$, 
amplitude of mass fluctuations, $\sigma_8 = 0.9$, and mean baryon density, $\Omega_{\rm b} = 0.04$. 
The size of the periodic simulation box is $35.325 h^{-1}$Mpc and the initial redshift 
is $z = 49$. 
We put the high-resolution dark matter particles and gas particles in 
the region where a halo with mass 
$M_{\rm vir} = 1.2 \times 10^{12} h^{-1} M_{\odot}$ forms at $z = 0$.  
The region external to this is populated with high-mass dark 
matter particles (boundary particles), the function of which is to 
reproduce the appropriate tidal fields. 
The circular velocity, spin parameter and collapse redshift of the 
selected halo are $v_{\rm c}(r_{\rm vir}) = 155$ km s$^{-1}$, $\lambda
\equiv J |E|^{1/2}/(G M^{5/2}) = 0.038$ and $z_{\rm c} \simeq 1.5$,
respectively, where $z_{\rm c}$ is defined as the redshift at which
the main progenitor had half the final halo mass.

The masses of the SPH and high-resolution dark matter particles 
are $\sim 2.6 \times 10^6$ and $\sim 1.7 \times 10^7 h^{-1} M_{\odot}$, 
respectively. 
Thus the $N_{\rm th} = 128$ implies that halos larger than 
$\sim 2.6 \times 10^9 h^{-1} M_{\odot}$ are allowed to have BHs. 
The gravitational softening length are kept fixed in comoving coordinates 
for $z > 3$; thereafter they are frozen in physical units at a value 
(of the equivalent Plummer softening) of $\epsilon = 0.5$ and $1$ kpc 
for the baryonic particles (SPH particles, stars, and BHs) and high-resolution 
dark matter particles, respectively. The gravitational force obeys the exact 
$r^{-2}$ law at $r > 2.8 \epsilon$. 

\subsection{Black hole particles}

When we find a halo with $N_{\rm dm} \ge N_{\rm th}$ that does not contain 
any BH particles and is not contaminated by the boundary particles,  
we turn the star particle that has the largest 
stellar density into a seed BH.
We thus discriminate the BH mass from the mass of the BH particle 
(particle mass). 
The former is updated every time-step according to the mass 
accretion rate obtained from the radiation drag model. 
Once BH mass exceeds its particle mass, the BH particle increases 
its particle mass by absorbing gas particles $n$ in the star-forming region 
around it with the probability:
\begin{equation}
p_n = \frac{M_{\rm BH} - M_{\rm p}}{N_{\rm SFR}} \frac{1}{M_n}, 
\end{equation}
where $M_{\rm BH}$ and $M_{\rm p}$ are, respectively,  the BH mass and the mass 
of the BH particle, $N_{\rm SFR}$ is the number of gas particles in 
the star-forming region, and $M_n$ is the mass of the gas particle $n$.

BH particles can increase their BH masses and particle masses by BH mergers.
However, how long it would take to harden a BH binary until finally gravitational wave 
emission becomes important is still unclear \citep{mf04, esc04}.
We simply assume that BHs can merge into a single BH if two or more BHs 
are within a gravitational softening radius and they are gravitationally bound.  
Therefore the BH merger rates obtained in our simulations are likely to be the 
maximum possible rate. 
Technically, we apply the friends-of-friends algorithm to find groups of BHs which 
may potentially merge originating new BHs. 

An interesting problem arises owing to the finite numerical resolution. 
In reality, supermassive BHs are much heavier than stars and 
dark matter. 
Therefore they are quickly brought to the galactic centres and stay at the 
bottom of the potential wells owing to the dynamical friction. 
In cosmological simulations, the mass of a BH particle is comparable to 
those of gas and stellar particles and lighter than the dark matter 
particles in the early stage of its evolution. 
As a result, BH particles tend to oscillate around galactic centres with 
the velocity dispersions of the stellar systems. 
To circumvent this problem, we allow BH particles to behave as peak tracers at 
galactic centres. 
At each time-step, we compute local stellar density fields within the 
gravitational softening radii of BH particles, $\epsilon$. 
We then displace the BH particles by a small distance, $\Delta l$, 
along the local density gradients. 
After several tests, we chose  
\begin{equation}
\Delta l$ = ${\rm MIN}(0.01\epsilon, 0.03 \  |\bm{v}| \ \Delta t), 
\label{equation:correction}
\end{equation}
where $\bm{v}$ is the velocity of a BH particle and $\Delta t$ is the time-step. 
This $\Delta l$ correction sufficiently suppresses the oscillation of the BH particles 
at galactic centres. 

We adopt a maximum time step for BHs, $\Delta t_{\rm max} = 1$ Myr, in order to follow mass accretion and associated 
feedback with a reasonable time resolution. Hence a time step for a BH particle is defined as 
$\Delta t_{\rm BH} = \min(\Delta t_{\rm max}, \Delta t_{\rm dyn})$, where $\Delta t_{\rm dyn}$ is a time step 
determined by the dynamics. 

\subsection{Models}

We first compare three simulations in order to show the effects of AGN feedback. 
The first simulation does not have feedback from AGN. We refer this model as the `no-fb' 
model. In the second model, we distribute 1\% of jet energy to the ambient halo gas particles 
defined by their densities, $\rho < 0.1 \rho_{\rm th}$ in the form of thermal energy using SPH smoothing.
Since only 1\% of jet energy is available for heating, we call this model the `weak-fb' 
model. In the `reference' model, we adopt the same AGN feedback scheme as in the weak-fb model but 
assume 10\% of jet energy is available to heat the ambient halo gas.
We will explore other models in which we change the implementation of AGN feedback and 
strength of SN feedback in \S\ref{PARAMETERS}.

\section{Results}

\begin{figure*} 
\includegraphics[height=20cm]{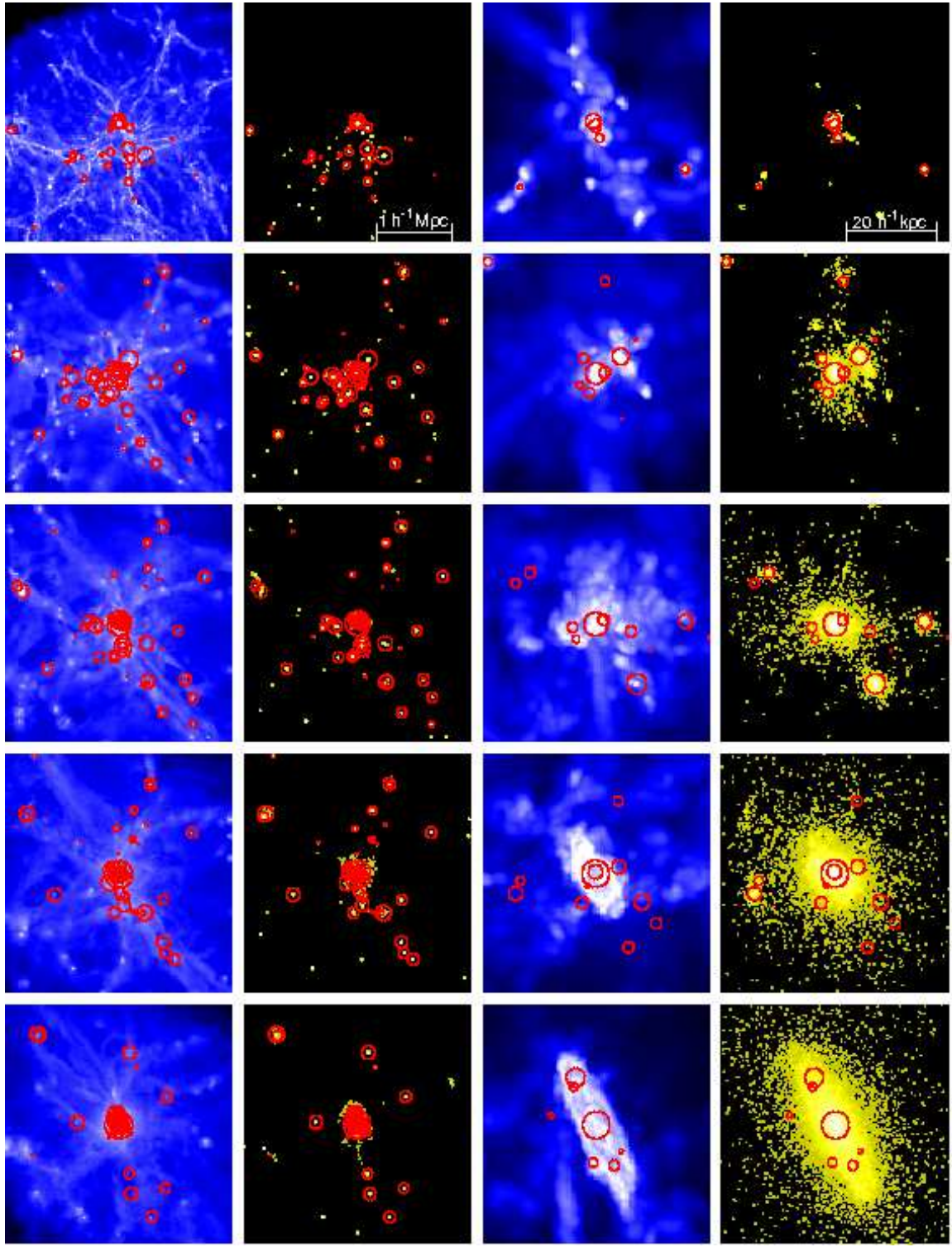} 
\caption{ 
Snapshots of the \sm{reference} model ($x$--$y$ projections).  
The rows correspond to $z = 6, 3, 2, 1,$ and $0$ from top to bottom. 
The first and the second columns show gas and stellar distributions in the comoving 3 $h^{-1}$ Mpc 
cubes centred on the main progenitor, respectively, where the brightness represents three dimensional 
density. 
The third and the fourth columns show gas and stellar distributions in the physical 50 $h^{-1}$ 
kpc cubes, respectively. 
The circles in each panel indicate the position of the BH particles and their radii are proportional to 
the log of the BH masses. 
}
\label{SNAPS}
\end{figure*} 

To give a visual impression of our simulations, we show some snapshots of the simulation 
with AGN-radio feedback (\sm{reference} model) in Fig.~\ref{SNAPS}. 
There are a number of merger events at high redshift ($z > 1$). 
The disk formation starts at $z \sim 1$, and finally a disk galaxy forms.  
A lot of BHs form in the small halos and they are brought to the main progenitor 
by merging satellites. 
The BH particles nicely trace the centres of the stellar objects 
resulting from our correction applied to the positions of BH particles. 
There are also many BHs which are not associated to any stellar objects. 
They are scattered into the halo when small satellites, 
which harbour a small BH at their centres, are destroyed during mergers. 
Most of the drifting BHs hence have very small mass. 

In the following sections,we present the main results of the paper. In \S5.1,
we compare the results of simulations run with the standard feedback parameters
(the `reference simulation') with simulations using weaker AGN feedback 
(the `weak-fb' model) and no AGN feedback at all (the `no-fb' model). 
In \S5.2 we look at the importance of the AGN feedback for the bolometric
luminosity of the forming galaxy, while in \S5.3, we explore the dependence of
the results on the way in which the feedback is coupled to the surrounding gas 
and the impact of assuming a lower efficiency of the feedback from SNe.

\subsection{Simulations with different AGN feedback efficiencies}

\begin{figure} 
\begin{center}
\includegraphics[width=8.3cm]{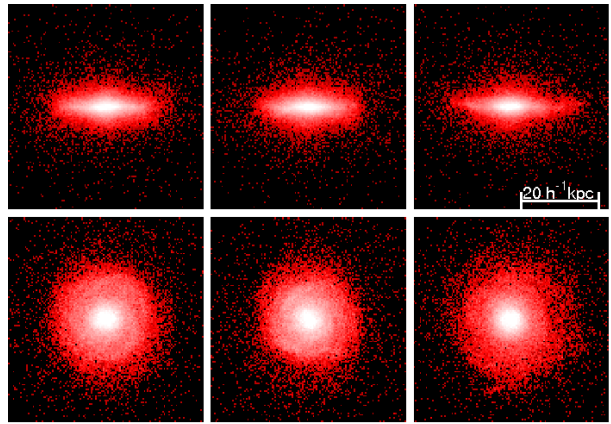} 
\end{center}
\caption{ 
Projected stellar distributions within 50 $h^{-1}$ kpc boxes centred on the galaxies 
at $z = 0$. 
The left, middle, and right panels show the \sm{no-fb}, \sm{weak-fb}, and \sm{reference} 
simulations, respectively. 
The viewing angles are chosen to be edge-on for upper panels and face-on 
for lower panels.  The brightness indicates the projected stellar mass density, 
and the same scaling is used for each simulation. 
}
\label{DISC}
%\end{figure*} 
\end{figure}

Now we compare galaxies formed in three simulations, i.e. the `no-fb', `weak-fb', and 
`reference' simulations. 
We first show the stellar distributions within $50 h^{-1}$ kpc 
boxes centred on the galaxies at $z = 0$ in Fig.~\ref{DISC}. 
The edge-on and face-on projections
are selected to be perpendicular and parallel to the angular momentum vector 
of the stellar component within the central $10 h^{-1}$ kpc sphere. 
In all of the models, the galaxy has an extended stellar disk. 
The \sm{no-fb} and \sm{weak-fb} models are very similar. 
From the face-on view, it is found that the \sm{reference} galaxy has a 
disk with lower surface stellar density. 
%
%\begin{figure*} 

\begin{figure} 
\begin{center}
\includegraphics[width=8.3cm]{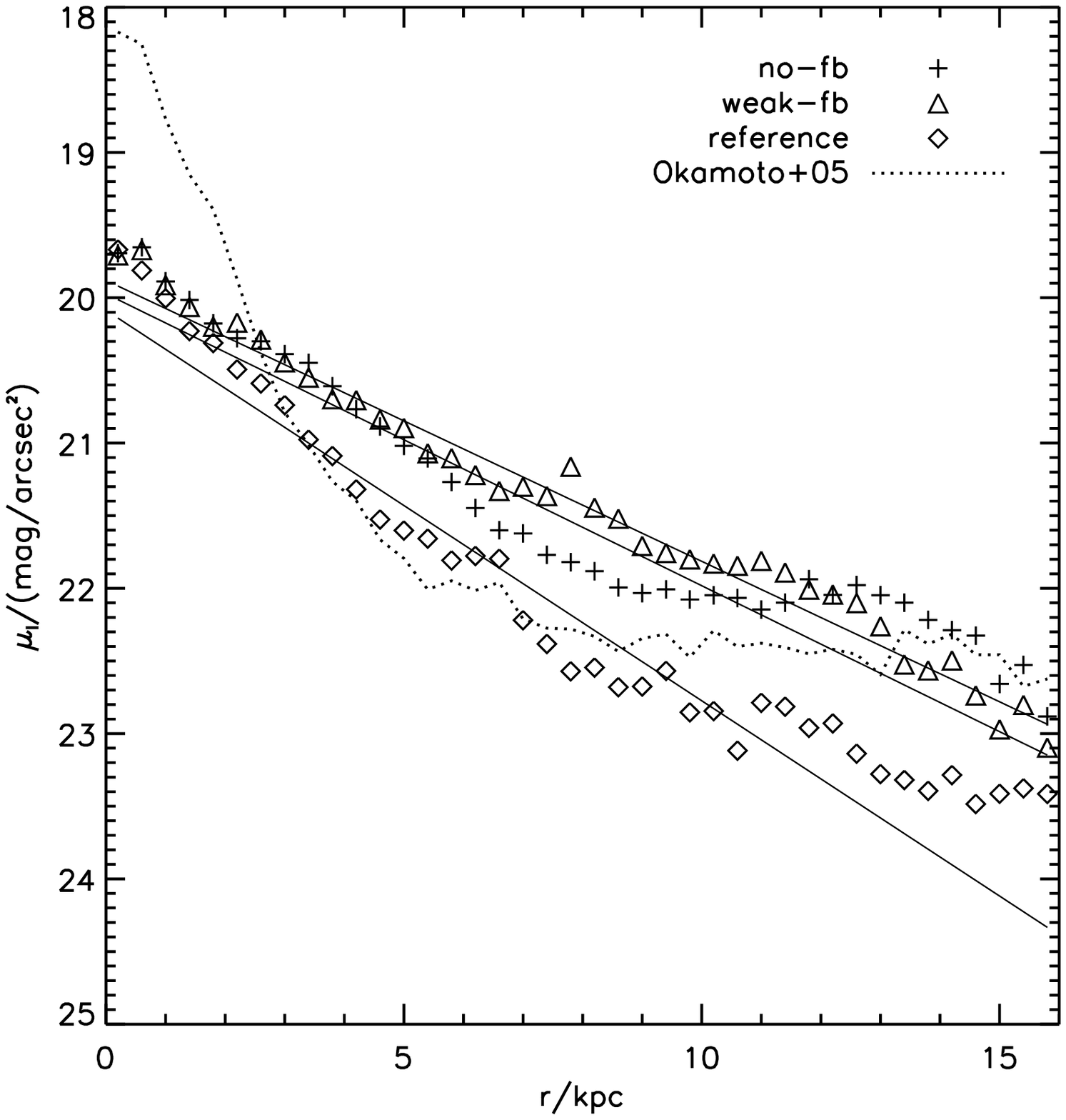} 
\end{center}
\caption{ 
  Surface brightness profiles in the $I$ band for the \sm{no-fb} (pluses), \sm{weak-fb} (asterisks), 
  \sm{reference} (diamonds) galaxies, and for the galaxy with the \sm{shock-burst} model of 
    Okamoto et al. (2005; dotted-line). 
    The solid lines are exponential fits for $r < 15$ kpc with the scale lengths
    5.4, 5.6, and 4.0 kpc for the \sm{no-fb}, \sm{weak-fb}, and \sm{reference} galaxies, 
  respectively. 
}
\label{SURFACE}
%\end{figure*} 
\end{figure} 
Fig.~\ref{SURFACE} shows the $I$ band surface brightness profiles for these 
galaxies. We also show the surface brightness profile of a disk galaxy 
obtained by \cite{oka05} using the same initial condition. 
An interesting difference from the surface brightness profile of the 
galaxy by \cite{oka05} can be seen at the centre. 
Our galaxies do not show the extreme central concentration.
The origin of this difference is the different implementation of SN feedback. 
While we implement it as kinetic winds, they deposited 
feedback energy in the form of thermal energy. 
Therefore they had to calculate the pressure gradients caused by the thermal 
feedback to blow the gas away. Unfortunately, the SPH does not allow 
steep gradients in physical quantities by definition unless one 
uses a sufficiently high resolution. 

All our three galaxies show similar central surface brightness. 
The main difference due to AGN feedback can be seen in the surface brightness 
of the disks.  
While \sm{no-fb} and \sm{weak-fb} galaxies show the almost same surface brightness 
profiles, the \sm{reference} model has a less extended disk. 
We do not try to determine the bulge-disk decomposition based on the surface brightness profiles since they provide only partial information
about the morphology of galaxies. Below, we will show the decomposition based 
on the dynamics of stars \citep{aba03b}. This is a more accurate way to 
characterise the relative importance of the bulge and disk components. 

\begin{figure} 
\begin{center}
\includegraphics[width=8.3cm]{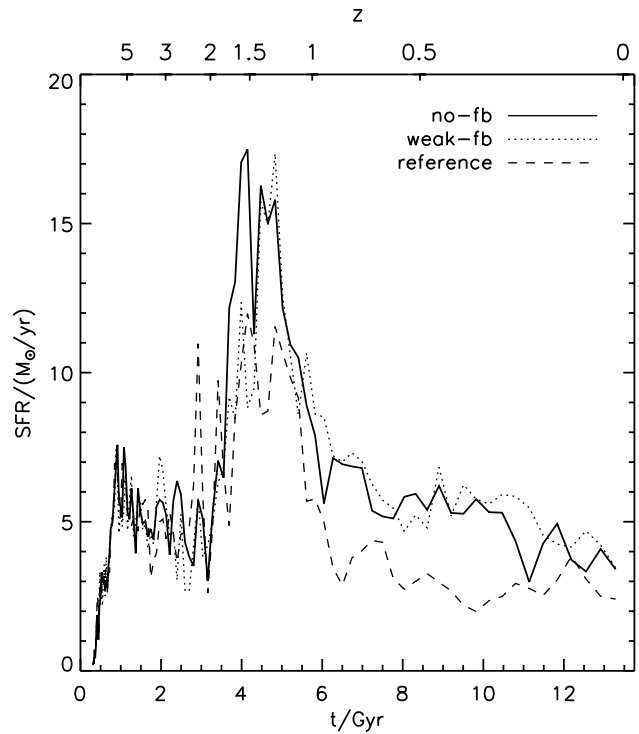} 
\end{center}
\caption{ The formation history (star formation rate as a function of time) of the stars 
    that lie within $25 h^{-1}$ kpc from the galactic centre at $z = 0$. 
    The solid and dotted lines correspond to the galaxies in the 
    simulation without and with BHs, respectively.  }
\label{SFH}
\end{figure} 
To investigate effects of AGN feedback quantitatively, we present the star 
formation histories of the stars that lie within $25 h^{-1}$ kpc from the 
galactic centres at $z = 0$ in Fig.~\ref{SFH}.  
All galaxies have qualitatively similar star formation histories and have peak 
star formation rates at $z \sim 1.5$. 
During the starburst ($t = 2.5 - 6$ Gyr), the \sm{reference} galaxy shows lower 
peak star formation rate compared with other galaxies.
As we will show later, the AGN is not a RIAF during the starburst. 
Hence AGN feedback from a RIAF is not responsible for this lower star formation rate. 
The feedback from standard disk (equation (\ref{eq:ljet_sdb})) might suppress the 
starburst. However we should not overinterpret the differences during the starburst
because the system is strongly non-linear during this epoch and hence small changes 
can result in relatively large difference. 
In fact, the model which employs stronger AGN feedback has a higher star formation 
rate during the starburst than in the \sm{reference} model as we will show later 
in Fig.~\ref{DIFFFB}. 

After the starburst, the \sm{reference} galaxy has significantly 
lower star formation rate than other galaxies. 
This suppression of star formation is likely to be caused by AGN feedback from a RIAF. 
The lower star formation rate in the \sm{reference} galaxy  explains its 
less extended disk in Fig.~\ref{SURFACE} because it is the star formation after 
$z \sim  1$ that builds up the stellar disk (see Fig.~\ref{SNAPS}). 
The \sm{weak-fb} galaxy has almost the same star formation rate as the \sm{no-fb} galaxy. 
Hence it can be said that, in order to give a visible impact on galaxy formation, 
at least $\sim 10\%$ of jet energy has to be given to the halo gas. 
We will later explore the case in which the jet energy is not given to the diffuse halo gas 
but delivered to the surrounding ISM. 

\begin{figure*} 
\begin{center}
\includegraphics[width=16cm]{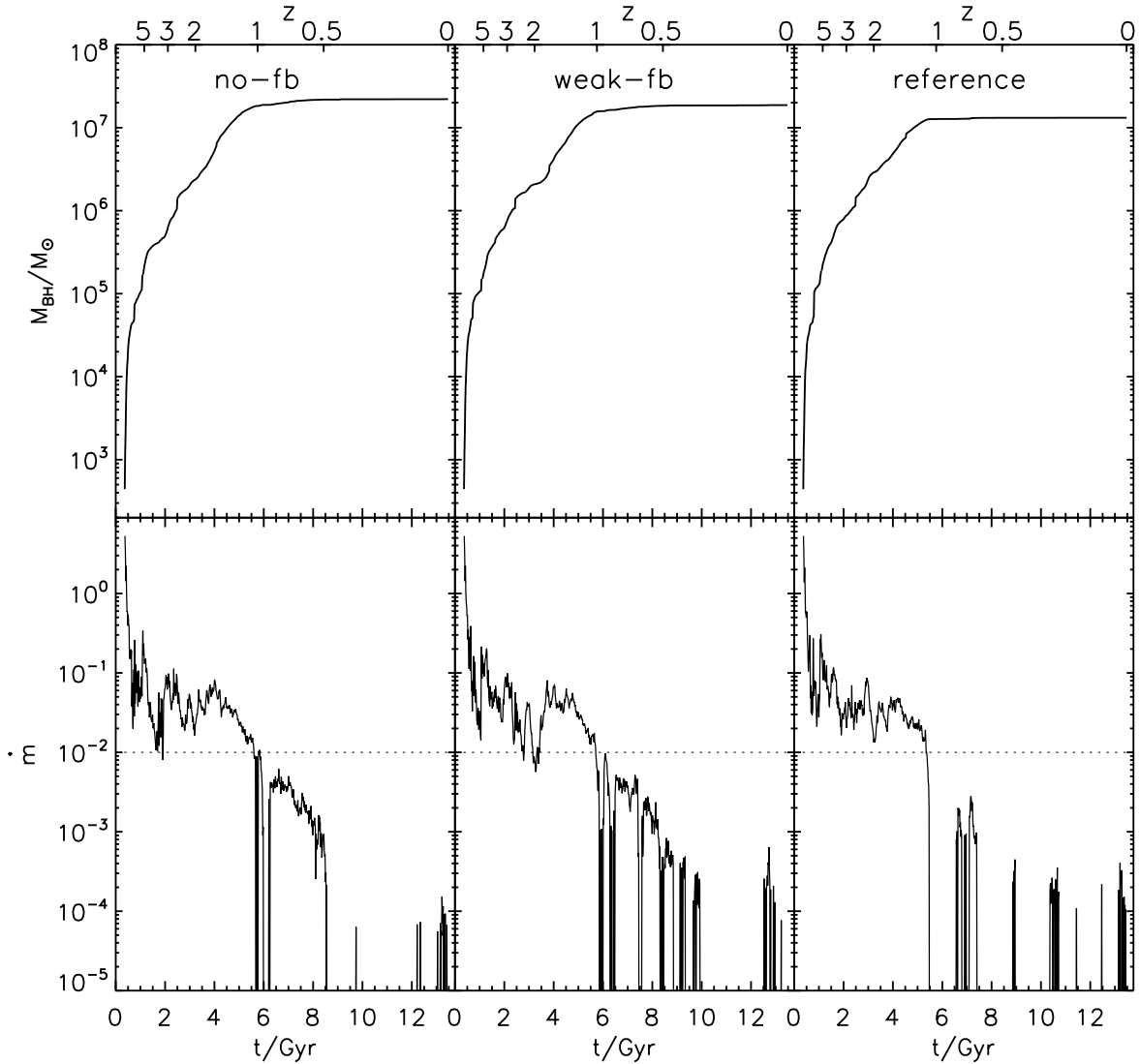} 
\end{center}
\caption{ 
  Upper panel: the mass growth of the central BH of the main progenitor. 
    Lower panel: the mass accretion rate onto the central BH of the main 
    progenitor in units of the Eddington rate 
    ($\dot{m} = \dot{M}/\dot{M}_{\rm Edd}$).
    The horizontal dashed line indicate the critical mass accretion rate 
    $\dot{m}_{\rm crit}$ below which the accretion flow becomes radiatively inefficient (RIAF). 
    The \sm{no-fb}, \sm{weak-fb}, and \sm{reference} simulations are shown from left to right
    panels. 
}
\label{MDOT}
\end{figure*} 
Since the strength of AGN feedback in our model is closely related to the central BH mass and 
the accretion rate onto the BH, 
we plot the evolution of the central BH of the main progenitor and
the mass accretion rate in units of the Eddington rate in Fig. \ref{MDOT} in each model. 
Note that the mass accretion rate $\dot{M}$ is different from the 
mass growth rate, $\dot{M}_{\rm BH}$; the latter includes the mass increase 
due to BH mergers. 
In all simulations, the central BHs rapidly evolve until $t \sim 5.5$ Gyr 
at which the starburst ceases. 
The final BH masses reach $\sim 10^7 M_{\odot}$. 
The stepwise jumps seen in the evolution of the BH mass indicate the mass growth 
by mergers. 

The consumption of the available gas for star formation explains the 
decline of the accretion rate onto the central BH.  
At $t \simeq 5.5$ Gyr $\dot{m}$ reaches the critical accretion rate $\dot{m}_{\rm crit}$, 
below which the accretion disk becomes a RIAF. 
Our model has the maximum AGN feedback when the mass accretion rate is equal 
to the critical accretion rate. This epoch is indicated by the crossing of the solid 
line and the horizontal dashed line in the lower panel of Fig.~\ref{MDOT}. 
Once the accretion disk becomes a RIAF after $z < 1$ , in the \sm{reference} 
model, the accretion rate drops to zero rapidly and there is no mass accretion 
after the strongest production of AGN feedback at $t \sim 5.5$ Gyr except for some episodic 
accretion events.
This implies that there is almost no star formation activity at the centre of the 
main progenitor at $z<1$ for this galaxy.  
On the other hand, lacking the strong AGN feedback, there are not such sudden decrease in 
accretion rates in the \sm{no-fb} and \sm{weak-fb} models. 
The gradual decline is caused by consumption of gas due to star formation and 
associated SN feedback(winds). 

Our implementation of AGN feedback cannot directly suppress the star formation 
because it injects energy not into the star forming dense gas but into the diffuse 
halo gas ($\rho < 0.1 \rho_{\rm th}$). Thus it is the suppression of the 
cooling of the halo gas around the centre of the galaxy which is responsible 
for the halting of star formation around the BH in the \sm{reference} simulation. 
The impulsive feature seen in the accretion rate implies the existence of a 
self-regulated cycle that operates between gas cooling, star formation, and 
production of jets in the RIAF.
The fact that the central surface brightness is almost the same in the three  
models in spite that the galaxy with AGN feedback hardly form stars at its 
centre after $z \sim 1$ suggests that the vast majority of stars at the centre have 
already been formed at $z \sim 1$ in all galaxies.  

We now return to the separation of the disk and bulge components.
As noted earlier, it is best to separate the bulge and disk components
using a dynamical decomposition \citep{aba03b}. 
For this, we first compute the angular momentum $J_z$ of 
each star particle parallel to the net angular momentum of stars 
within $10 h^{-1}$ kpc, and the angular momentum of the corotating 
circular orbit, $J_c(E)$, where $E$ is the energy of each particle. 
The ratio $J_z/J_c$ defines an orbital circularity. 
\begin{figure} 
\begin{center}
\includegraphics[width=8.3cm]{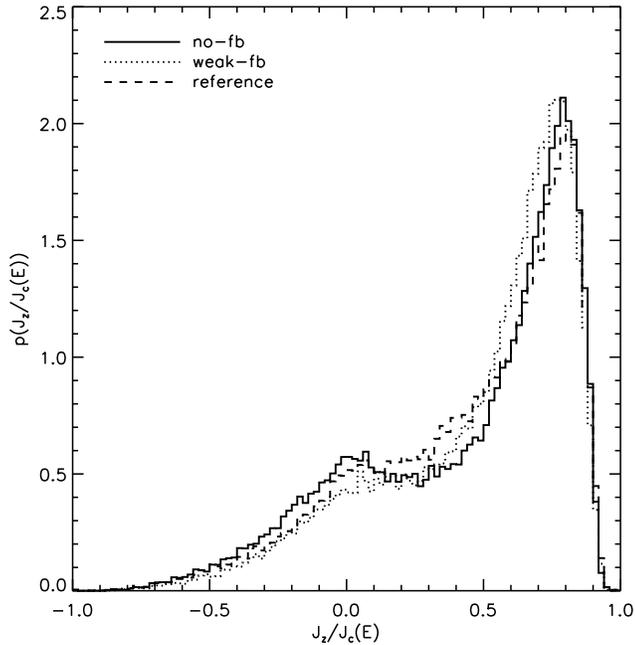} 
\end{center}
\caption{ 
  Mass-weighted probability distributions of the orbital circularity, 
$J_z/J_c(E)$, of stars within $25 h^{-1}$ kpc from the galactic centres. 
The solid, dotted, and dashed lines indicate the `no-fb', `weak-fb', 
and `reference' galaxies, respectively.  
}
\label{DECOMPOSITION}
\end{figure} 
In Fig.~\ref{DECOMPOSITION}, we show the probability distribution 
of this orbital circularity for stars within $25 h^{-1}$ kpc 
from the centre of each galaxy. A disk component should have 
$J_z/J_c(E) \simeq 1$ and we see that such a component clearly dominates  
in each galaxy. 
Comparing to the results by \cite{oka05}, the suppression of the formation 
of spheroidal components seems to rely largely on the strong SN feedback by 
winds. \cite{gov07} also produced disk dominated galaxies in halos 
with quieter merger histories (in which the epoch of the last major merger is $z_{\rm lmm} > 2$)
using a different feedback recipe from ours. 
To identify the stars in the spheroid, we assume a non-rotating spheroid, 
i.e. that stars in the spheroid are symmetrically distributed around zero 
in Fig.~\ref{DECOMPOSITION}, 
where all stars having $J_z/J_c(E) \le 0$ are identified as a half of the 
spheroidal component. 
Another half of the spheroid with $J_z/J_c(E) > 0$ are defined statistically 
so that the total angular momentum of the spheroid becomes zero. 
All remaining stars are identified as the disk component. 
As expected from the similarity in the star formation histories, 
all galaxies have similar distributions in this plane while the disc component 
in the \sm{reference} model is slightly less significant because of the relatively 
strong AGN feedback.

\begin{table*}
\caption{Total and spheroid's mass, $g$ and $r$ band total luminosities, 
  and disk-to-total mass and luminosity ratios 
  for the galaxies at $z=0$. }
\label{BT}
\begin{center}
\begin{tabular}{@{}lcccccccc}
\hline
  & $M_{\rm tot}/\ M_\odot$ &
  $M_{\rm SP}/\ M_\odot$ &
  $M_{g}$ &
  $M_{r}$ &
  $\left(\frac{D}{T}\right)_{\rm Mass}$ &
  $\left(\frac{D}{T}\right)_{g}$ &
  $\left(\frac{D}{T}\right)_{r}$ &\\
\hline 
no-fb & $6.01 \times 10^{10}$ & $1.98 \times 10^{10}$ & -20.8 & -21.3 & 0.67 & 0.83  & 0.79 \\
weak-fb & $5.98 \times 10^{10}$  & $1.44 \times 10^{10}$ & -20.9  & -21.3 & 0.76 & 0.88  & 0.85 \\
reference & $4.32 \times 10^{10}$  & $1.21 \times 10^{10}$ & -20.6  & -21.0 & 0.72 & 0.86  & 0.83 \\
\hline
\end{tabular}
\end{center}
\end{table*}
In Table~\ref{BT} we show the total and spheroid's masses, total luminosity, and 
the disk-to-total ratios in mass, $g$ band, and $r$ band. 
Since colour of a stellar population is sensitive to its metallicity, we here 
include the metallicity dependence to compute the luminosity at each passband by using 
P$\acute{\rm E}$GASE2 \citep{pegase}. 
The masses of the spheroidal components directly reflect amplitude of starbursts and 
therefore the \sm{no-fb} galaxy has the most massive spheroidal component. 
Nevertheless, all galaxies have disk-dominated morphology. 

\begin{figure} 
\begin{center}
\includegraphics[width=8cm]{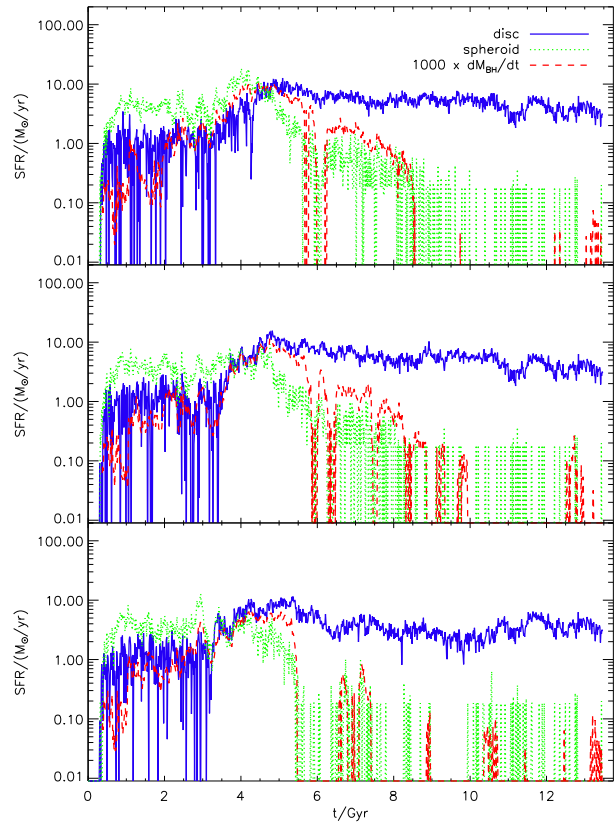} 
\end{center}
\caption{ 
  Star formation histories for the disk and spheroid stars within $25 h^{-1}$ kpc from the centre of each galaxy at $z = 0$. 
The \sm{no-fb}, \sm{weak-fb}, and \sm{reference} galaxies are shown in the top,
middle, and bottom panels, respectively. 
The star formation histories for the disk components are represented by the 
solid lines and those for the spheroidal components are indicated by the dotted lines. 
We also show the mass accretion rates, which are multiplied by 1000,  onto the central BHs in the main progenitors (dashed lines). 
Note that while star formation not only in the main progenitors but also in all other building blocks is counted 
in the star formation histories, the mass accretion rates only represent the rates onto the central BHs of the main progenitors. 
}
\label{SFHDECOUPLE}
\end{figure} 
It is interesting to see when each component forms and how the formation history of each component 
is related to the accretion history of the central BH. 
In Fig.~\ref{SFHDECOUPLE}, we show star formation histories of the disk and spheroidal components in each galaxy at 
$z = 0$. 
We also show the rescaled accretion rate onto the central BH of the main progenitor of each galaxy. 
Note that the star formation rates include star formation in all building blocks (all progenitors) but the mass 
accretion rates are only the rates onto the central BHs of the main progenitors. 
Thus at high redshift, the mass accretion rates presented in the figure considerably underestimate 
the net accretion rates. 
Since the galaxies do not undergo any significant mergers after $z \sim 1$, it is safe to compare 
the star formation rates and the mass accretion rates at low redshift. 
We find that there are good correlations between the star formation rates of the spheroid and the mass 
accretion rates onto the central BHs at low redshift, in particular between $t \sim 5.5$ and 8 Gyr.  
After the starburst, the star formation rates of the spheroids rapidly decreases in all simulations. 
The accretion rates onto the BHs also decrease at the same time. 
The contribution from disk stars to the mass accretion rate due to the radiation drag effect should be 
largest during the peak of star formation in the disc component ($t \simeq 5$ Gyr) because at 
this epoch the disc is gas rich and very compact. We have however confirmed that even at this epoch 
the contribution from disc stars is less than 20\%  and most of the time less than 5\% in all models. 
Considering the fact that half of the spheroid stars are defined {\it statistically}, the correlation between 
the star formation rate and the mass accretion rate of the spheroid in each galaxy is significant. 
After $t \sim 8$ Gyr there are many short lived star formation episodes in 
the spheroids, which do not correspond to gas accretion events. 
These star formation events occur in small satellites before they become a part 
of the spheroidal components by minor mergers.

\begin{figure} 
\begin{center}
\includegraphics[width=8cm]{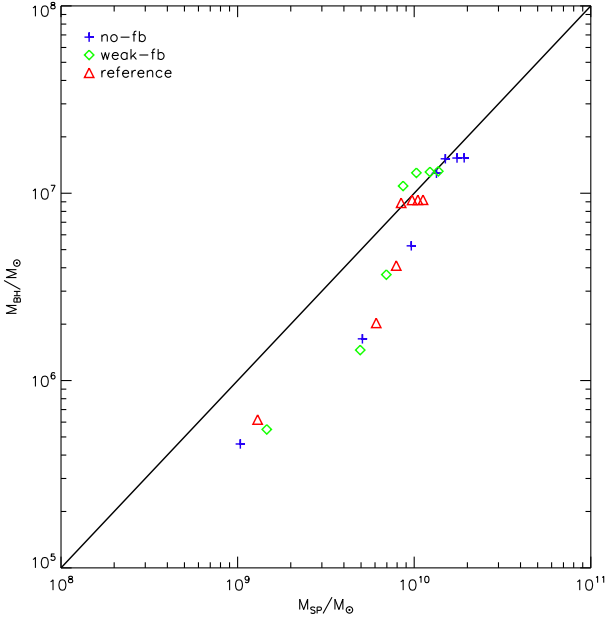} 
\end{center}
\caption{ 
The relation between mass of the central BHs of the main progenitors and the stellar mass of 
the host spheroids at $z = 3$, 2, 1.5, 1, 0.5, 0.2, and 0. 
The relations for the \sm{no-fb}, \sm{weak-fb}, and \sm{reference} simulation are shown 
by the plus signs, diamonds, and triangles respectively. 
The solid line indicates $M_{\rm BH} = 0.001 M_{\rm SP}$. 
}
\label{MM}
\end{figure} 
Next we show in Fig.~\ref{MM} the evolution of the relations between mass of the central BHs of 
the main progenitors and the mass of the host spheroids at $z = 3$, 2, 1.5, 1, 0.5, 0.2, and 0. 
We here define the main progenitors as stars within 10\% of the virial radii at each redshift. 
We have confirmed that the spheres with radii of $0.1 r_{\rm vir}$ contain the stellar systems defined 
as main progenitors by the friends-of-friends algorithm in \S \ref{RelativeImportance}.  
We have also checked by eye that at selected redshifts the main progenitors are not undergoing major 
mergers and the spheres do not contain massive satellites, both of which ruin the dynamical decomposition.
We then compute the mass of the spheroidal component of each main progenitor based on the dynamical 
decomposition, noting that this method of decomposition is less vulnerable to insufficient resolution 
compared with the decomposition based on surface density profiles. 
This feature is important at high redshift where the size of the galaxies are 
much smaller than that at $z = 0$. 
Fig.~\ref{MM} shows that the $M_{\rm BH}$-$M_{\rm SP}$ relations evolve similarly in all simulations
because the BHs and spheroids mainly increase their mass during the starbursts and 
our AGN feedback is not effective during this phase. 
Before the starburst, the BH mass is far below the local relation, $M_{\rm BH} \sim 10^{-3} M_{\rm SP}$, 
and then it rapidly moves towards the local relation during the starburst ($z = 2 \sim 1$). 
The delay in the mass evolution of the BHs compared to the spheroids is resulting from the fact 
that the mass accretion rate due to the radiation drag model depends on the 
metallicity of the ISM. The mass accretion by this model is not efficient until the ISM is 
metal enriched and the optical depth becomes $\sim 1$.  
After the starburst, the mass of BHs stays almost constant as seen in Fig.~\ref{MDOT}. 
The relation is imprinted in the radiation drag model because the model provides a good 
correlation between star formation in a spheroid and the mass accretion 
onto a BH.  
It is important to note that the radiation drag model predicts considerably lower mass ratio between central 
BHs and their host spheroids before the starburst.

\begin{figure} 
\begin{center}
\includegraphics[width=8.3cm]{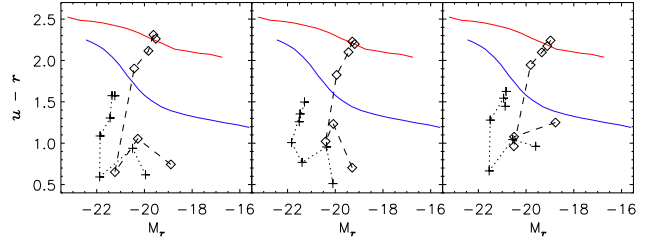} 
\end{center}
\caption{ 
  Colour-magnitude diagrams for the `no-fb' (left), `weak-fb' (middle), and 
  `reference' (right) galaxies. 
  The galaxies at $z = 3, \ 2, 1.5, 1, 0.5, 0.2$ and $0$ are indicated by the plus 
  signs connected by the dotted line. 
  The colours of their spheroidal components are also indicated by the diamonds. 
  The upper and lower thin solid lines represent mean colours of the red and blue 
  populations taken from Baldry et al. (2004). 
  The $u - r$ colour of simulated galaxies are converted to the SDSS colours by using 
  $(u-r)$(SDSS) = $(u - r)$(AB) + 0.05 (Abazajian et al. 2003) to compare with the data by 
  Baldry et al. (2004). 
}
\label{COLOUR}
\end{figure} 
In order to illustrate how the colours of our galaxies evolve, we show the evolutions of 
the main progenitor of each galaxy and its spheroid in a colour-magnitude diagram (Fig.~\ref{COLOUR}). 
The mean colours of the red and blue populations of the SDSS galaxies \citep{baldry04} 
are indicated by solid lines. 
All galaxies show similar evolution on the colour-magnitude plane. This confirms that the 
AGN feedback has only minor effects on evolution of disk galaxies. 
The colour of the galaxies and their spheroids at $z = 0$ are consistent with the observed 
colours of local blue and red galaxies, respectively. 

\subsection{Relative importance of the AGN} \label{RelativeImportance}

\begin{figure*} 
\begin{center}
\includegraphics[width=16cm]{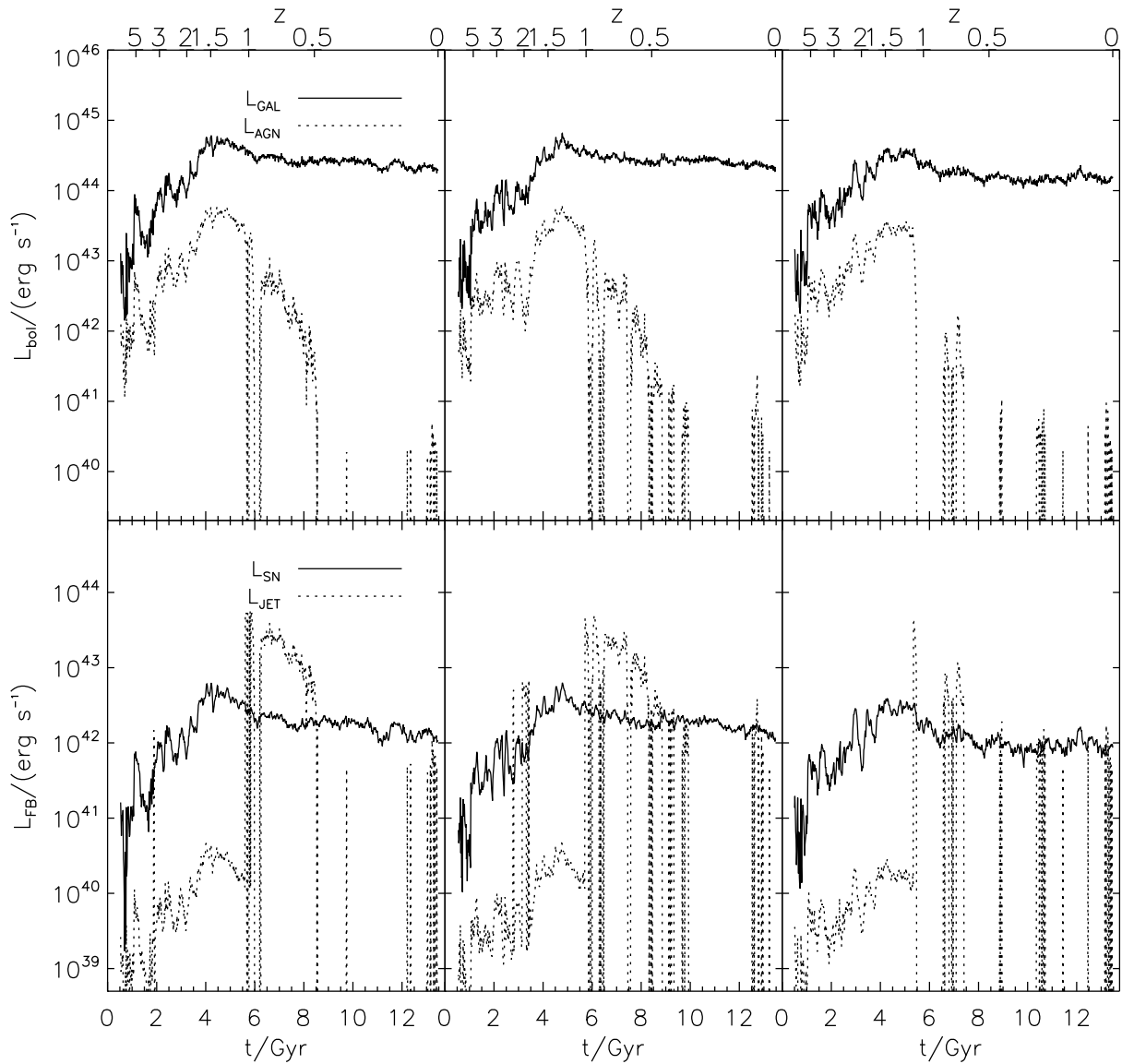} 
\end{center}
\caption{ Upper panels: Bolometric luminosities radiated by the stars in the 
  main progenitors (solid lines) and by the 
  AGNs of the main progenitors (dotted lines). 
  A 10\% radiative efficiency is assumed for standard thin disks and 
  the efficiency given by eq.~(\ref{RADEFF}) is assumed for RIAFs. 
  Lower panel: Raw feedback luminosities ejected by SNe (II and Ia) in 
  the main progenitors (solid lines)  and by jets from the central 
  BHs (AGNs) of the main progenitors (dotted lines). 
  From left to right, the \sm{no-fb}, \sm{weak-fb}, and \sm{reference} 
  simulations are shown. 
  %Note that feedback efficiencies, $\eta_{SN}$ and $\eta_{jet}$, are not multiplied. 
}
\label{RELATIVE}
\end{figure*} 
In order to show the relative importance of the AGN in 
galaxy formation, we show the bolometric luminosities generated by stars in 
the main progenitors and the bolometric luminosities radiated by the AGNs 
of the main progenitors in the upper panels of Fig.~\ref{RELATIVE} for our three 
simulated galaxies. 
A main progenitor is defined by applying the friends-of-friends algorithm 
for stellar particles with a linking length $l_{\rm link} = 0.02 \left\langle l \right\rangle$, 
where $\left\langle l \right\rangle$ is the the mean separation of dark matter particles. 
To estimate the luminosity of an AGN, we assume that the efficiency of conversion of rest mass energy 
of the accretion flow into radiation ($L_{\rm AGN}/(\dot{M} c^2)$) is $\eta_{\rm SD}=10 \%$ for 
a standard thin disk and for a RIAF we use the equation
\begin{equation} \label{eq:etariaf}
\eta_{\rm RIAF} = 0.1 \left( \frac{\dot{m}}{\dot{m}_{\rm crit}} + \delta + 10^{-3} \right) 
  \label{RADEFF}
\end{equation}
where $\delta$ represents the fraction of the 
turbulent energy which heats the electrons \citep{qua01}.
While the value of $\delta$ is quite uncertain, 
we employ $\delta = 0.1$ as commonly assumed in spectral models of RIAFs \citep{nem06}. 
Note that the radiative efficiency for a thin disk can be a factor of 
a few higher than $0.1$ in the case of a spinning BH, and may reach efficiencies as high as 
$42\%$ for an extreme Kerr BH \citep{nt73}. 

It is found that the main progenitors are always brighter than the AGNs for these 
galaxies as we expect for disk-dominated galaxies. 
Since the radiation drag is more efficient for more compact star forming region, we expect 
that the significance of the AGN luminosity would increase for galaxies which are undergoing 
gas-rich mergers. 
There is a notable difference 
among our models when the accretion flows become RIAFs after $z \sim 1$.
While all the AGNs have much lower luminosities than their host galaxies at low redshift 
because of the low accretion rates and the low radiative efficiency of the RIAF, 
only the AGN in the \sm{reference} galaxy has almost zero luminosity right after the accretion disc switches to a RIAF. In this simulation, the low accretion rate, due to the 
suppression of star formation around the BH by AGN feedback, makes the AGN invisible most 
of the time.  
In the \sm{no-fb} and \sm{weak-fb} simulations, the AGNs still continuously emit light 
for a few giga years  after the onset of the RIAFs. 
This result suggests that if 10\% of the jet luminosity is used to heat the ambient halo gas 
AGNs become quite dim and can be hardly observed after the starburst regime. 
It should be noted that it is beyond the scope of this work to incorporate 
in the simulations possible inclination effects in the appearance of the AGN, 
caused by dust obscuration inside the nucleus as in the obscuring dust torus 
model invoked in unified models of AGN (e.g. \citealt{anton93, krolik99}).

In the lower panels of Fig.~\ref{RELATIVE} we show the raw feedback luminosities 
ejected by the total SNe in the main progenitors and jet luminosities produced by the AGNs 
computed by using equations (\ref{SD}) and (\ref{RIAF}).  
Until the end of the starburst, the feedback energy is dominated 
by energy produced by SNe except for a few temporary RIAF modes seen in the \sm{weak-fb} 
simulation. 
The jet luminosities suddenly exceed the SN luminosities when the accretion flow becomes radiatively inefficient. 
Since in the \sm{no-fb} and \sm{weak-fb} simulations, the efficiencies, $\eta_{\rm AGN} $ by which the AGN jet 
energy is given to the surrounding halo gas are small ($\eta_{\rm AGN} = 0$ and 0.01 respectively), the effects of 
the AGN feedback are none or negligible in these simulations. 
In the \sm{reference} simulation, the efficiency, $\eta_{\rm AGN} = 0.1$ is sufficiently high to suppress 
the star formation around the BH therefore quenching the mass accretion to the BH. 
This is an important feature of our AGN feedback. 
The impulsive events in Fig. \ref{RELATIVE} represent epochs where the AGN would be seen as 
radio-loud, and they constitute the {\it radio-loud mode} of the AGN. 
The duration in which the AGN is in the {\it radio-loud mode} is strongly 
dependent on the feedback efficiency, since the AGN feedback terminates the {\it radio-loud mode} through the suppression of gas cooling and hence star formation.
Interestingly enough, in the \sm{reference} simulation, the AGN spends most of the time in the 
{\it radio-quiet mode} and only a fraction 8.8\% of the time in the {\it radio-loud mode}. 
Contrarily, in the \sm{no-fb} and \sm{weak-fb} galaxies, the AGNs spend $\sim 25\%$ of the time in 
the {\it radio-loud mode}. 
Note that we are likely to overestimate the star formation rate at the centre and hence AGN activity
particularly in the quiescent star forming regime because of the angular momentum transfer due to the artificial
and numerical viscosities which can bring gas particles to the centre \citep{oka03, kau07}.

\subsection{Dependence on feedback parameters and implementations of feedback}
\label{PARAMETERS}

In \S5.1, we compared simulations with different AGN feedback efficiencies. 
In this sections, we 
compare simulations which all have strong AGN feedback, but which differ in other aspects.
We compare three additional models. The models investigate the coupling of the AGN feedback,
the role of BH spin and the effect of the wind speed generated by SN feedback.

In the first model, AGN feedback 
energy is not given to the ambient halo gas (diffuse gas) but deposited to the 
nearest 40 ISM particles (dense gas). We use $\eta_{\rm AGN} = 1$ so that 
all the jet energy is used to heat the surrounding ISM because a model with 
$\eta_{\rm AGN} = 0.1$ does not show any visible difference from the \sm{no-fb} model. 
We call this model the \sm{ism-fb} model. 
In the second model, we assume the higher spin, $j = 0.9$, and the 
maximum feedback efficiency, $\eta_{\rm AGN} = 1$, as we may need such strong 
feedback in order to explain the X-ray properties of clusters of galaxies 
\citep[e.g.][]{nem07}. 
As a result, the efficiency of AGN feedback in this model becomes 17 times 
higher than in the \sm{reference} model .
We dub this model the \sm{max-fb} model.
In the third model, we use the same AGN feedback as in the \sm{reference} model 
but employ a slow wind speed, 250 km s$^{-1}$ for SN feedback in order to explore  
effects of the wind speed. We refer this model as the \sm{slow-wind} model. 

\begin{figure} 
\begin{center}
\includegraphics[width=8.3cm]{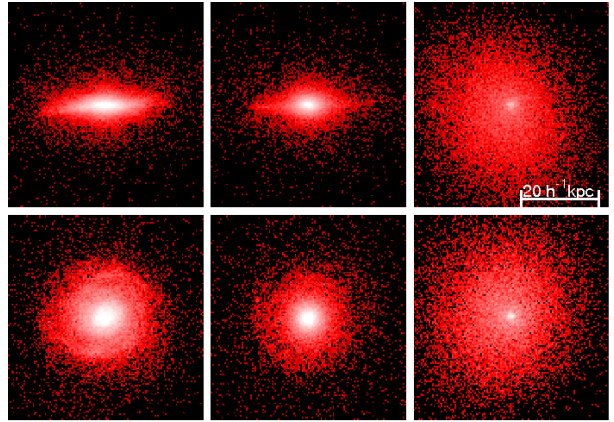} 
\end{center}
\caption{ 
Same as Fig.~\ref{DISC} but for the \sm{ism-fb} (left), \sm{max-fb} (middle), and 
  \sm{slow-wind} (left) models. 
}
\label{DISC2}
\end{figure}
In Fig.~\ref{DISC2}, we show the edge-on and face-on views of the stellar distributions 
of the galaxies produced by the three additional models. 
The \sm{ism-fb} model is similar to the \sm{reference} model. 
As expected the disk component of the \sm{max-fb} galaxy is less significant than 
that of the \sm{reference} model since stronger AGN feedback suppresses the disk 
formation more strongly. 
The \sm{slow-wind} model exhibits completely different morphology from others.  
There is no sign of a disk component. 

\begin{figure*} 
\begin{center}
\includegraphics[width=16cm]{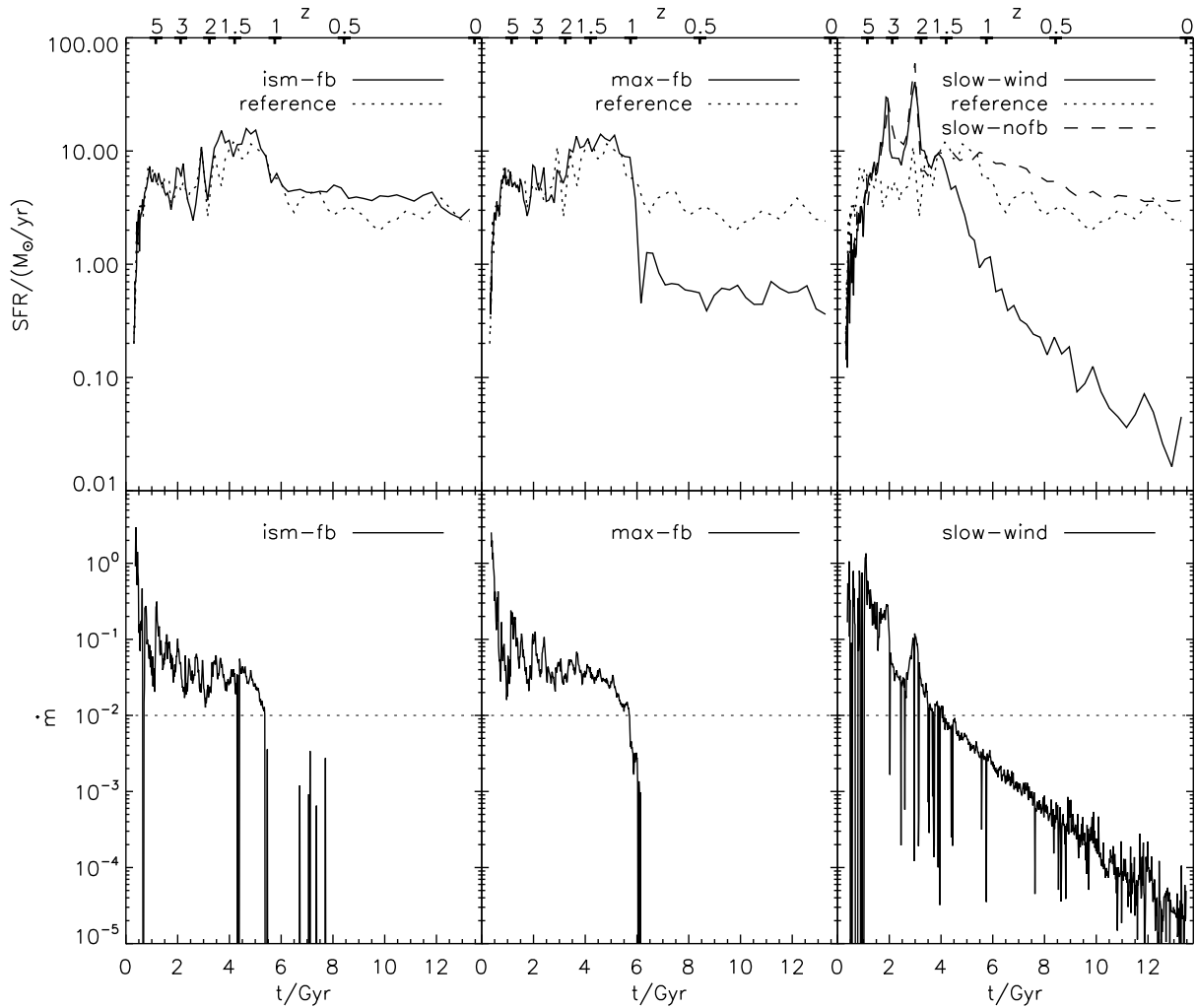} 
\end{center}
\caption{ 
  Upper panel: star formation histories for 
  the \sm{ism-fb} (left), \sm{max-fb} (middle), and \sm{slow-wind} (right) 
  models. The star formation history for the reference model (dashed line in Fig.~ref{SFH})
  is indicated by the dotted line. In the right panel, we also show the \sm{slow-nofb} model
  in which we use the same wind speed as in the \sm{slow-wind} model and do not include AGN 
  feedback (dashed line) for comparison. 
  Lower panel: same as the lower panel of Fig.~\ref{MDOT} but for the \sm{ism-fb} (left), 
  \sm{max-fb} (middle), and \sm{slow-wind} (right) models. 
}
\label{DIFFFB}
\end{figure*}
Now we carry out quantitative comparison by showing the star formation histories 
of the galaxies and the accretion histories to the central BHs (Fig.~\ref{DIFFFB}). 
In spite of the fact that we employ an order of magnitude higher 
AGN feedback efficiency in \sm{ism-fb} model, the star formation history 
of this galaxy is very similar to that of the \sm{reference} model. 
It suggests that delivering  AGN feedback energy to the diffuse hot gas causes  
the stronger feedback effect than depositing it into the surrounding dense ISM 
where cooling time is very short.  
The \sm{max-fb} galaxy shows a resembling star formation history to the \sm{reference} 
model (upper-middle panel) at high redshift ($t < 6$ Gyr) where the AGN harbours a thin accretion disc. 
Once the accretion flow becomes a RIAF, strong AGN feedback suppresses the star formation dramatically and 
this galaxy have far less star formation after $t \sim 6$ Gyr than the \sm{reference} galaxy. 
The \sm{slow-wind} galaxy has a completely different star formation history from others. 
Owing to the slow wind speed, 
it has a higher star formation rate at high redshift including two burst events 
at $t \sim 2$ and $3$ Gyr. The AGN in this galaxy becomes underfed earlier than in other models 
(at $t \sim 4$ Gyr) and the AGN feedback strongly suppresses the subsequent star formation.  
We also show the model which employs the same wind speed as the \sm{slow-wind} model 
but without AGN feedback (dashed line in the upper-right panel). 
By comparing these two slow-wind models, we can see that the AGN feedback is responsible 
for the suppression of the star formation at low redshift. The bulge-to-total mass ratio of 
the \sm{slow-wind} model is 0.9 and the $u - r$ colour of this galaxy is consistent with 
the mean colour of the red population at $z = 0$.

The evolution of the accretion rates in the \sm{ism-fb} and \sm{max-fb} models are similar to that 
in the \sm{reference} model particularly at high redshift ($t < 6$ Gyr) where the accretion disc is thin 
($\dot{m} > 10^{-2}$). 
In these two models, the accretion flows become RIAFs after the starbursts as in the \sm{reference} model. 
After that the mass accretion onto the central BH in the \sm{ism-fb} model is more strongly suppressed 
than in the \sm{reference} model, while the \sm{ism-fb} model has a higher star formation rate. 
Since the AGN feedback energy is deposited into the surrounding star forming ISM particles in this model, 
it can terminate star formation around the BH more directly than in the \sm{reference} model where 
the AGN feedback energy is distributed to the halo gas. 
In the {max-fb} model, there is no accretion event after $t \simeq 6$ Gyr. The strong AGN feedback in 
this model removes all the low angular momentum gas around the BH and prohibits subsequent star formation 
around the galactic centre. 
The \sm{slow-wind}  model shows a different accretion history from other models. 
Since the starbursts occur earlier in this model owing to the slow wind speed, the AGN also becomes underfed  
earlier ($t \sim 4$ Gyr) than in other models. 
Contrary to other models, the mass accretion continues at $t \le 4$ Gyr. 
This is because the slow wind speed, $250$ km $s^{-1}$, does not allow gas particles to escape 
from galactic centre and the AGN feedback energy is not given to the star forming gas. 
The accretion rate might become more episodic if we deposit a significant fraction of the AGN feedback energy 
to the central star forming gas as in the \sm{ism-fb} model. 

\begin{figure} 
\begin{center}
\includegraphics[width=8cm]{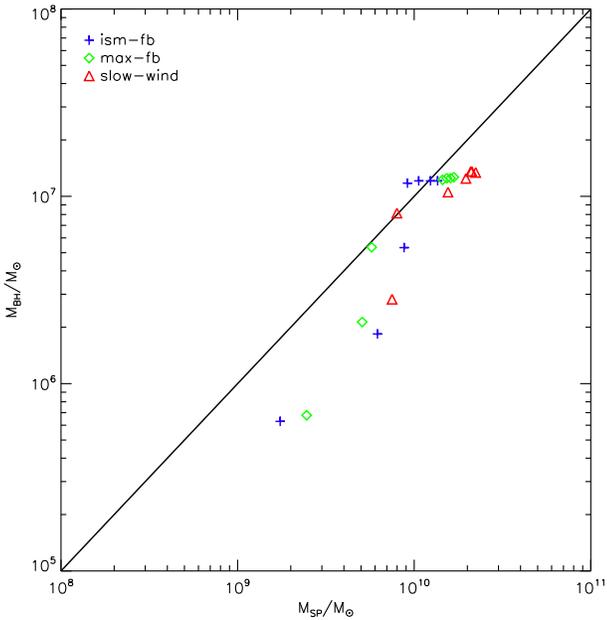} 
\end{center}
\caption{ 
Same as Fig.~\ref{MM} but for the \sm{ism-fb} (pluses), \sm{max-fb} (diamonds), and 
\sm{slow-wind} (triangles) models. 
}
\label{MM2}
\end{figure}
In Fig.~\ref{MM2}, we show the relations between mass of the central BHs of the main progenitors 
and the mass of the host spheroids at $z = 3$, 2, 1.5, 1, 0.5, 0.2 and 0.  
The relations for \sm{ism-fb} and \sm{max-fb} models evolve similarly to the relations 
for the models presented in Fig.~\ref{MM}. 
The relation for the \sm{slow-wind} model reaches the local relation earlier (at $z = 2$) 
  because this galaxy undergoes starbursts earlier than other galaxies. 
At low redshift, this model has slightly smaller BH mass to spheroid mass ratio 
than in other models. 
This is because infalling satellites increase the mass of the spheroid without changing the 
mass of the central BH very much when these infalling galaxies are disk dominated. 
Due to the slow wind speed, these infalling systems have more stars than in other models. 
Fig.~\ref{MM} and Fig.~\ref{MM2} tell us an important feature of the radiation drag model, that is, 
the $M_{\rm BH}-M_{\rm SP}$ relation after the last starburst does not depend 
on either the wind speed or efficiency of AGN feedback. 
In all models, the central BH masses are $\sim 0.1$\% of the mass of their host spheroids, 
$M_{\rm BH} \simeq 10^{-3} M_{\rm SP}$, when the accretion becomes radiatively inefficient. 

\section{Comparison with observations of AGN in spiral galaxies} \label{obs}

%X typical Mdot and Lbol during evolution
%X stochastic episodic activity, alternates between normal and active 
%X classify object as low-luminosity AGN, Seyfert - RL/RQ
As we have shown in \S\ref{RelativeImportance}, for $z \lesssim 1$ the bolometric luminosity of the AGN is feeble compared to the host galaxy and the nucleus undergoes several recurring episodes of activity. During these episodes when the nucleus is active, its characteristic accretion rate is in the range $\dot{m} \approx 10^{-4}-10^{-3}$, while the typical nuclear bolometric luminosities and jet powers are $L_{\rm bol} \approx 10^{40}-10^{42} \; {\rm erg}\ {\rm s}^{-1}$ and $L_{\rm jet} \approx 10^{42}-10^{43} \; {\rm erg}\ {\rm s}^{-1}$ respectively. Given these properties, at $z \lesssim 1$ during which the galactic disk is formed, the AGN can be considered to be in a highly sub-Eddington, low-luminosity or Seyfert state which undergoes a stochastic cycle of activity. The Seyfert nucleus is sometimes turned on, but most of the time is \qm{off} or inactive with $\dot{m} \approx 0$, resembling the nucleus of a normal spiral galaxy. The \qm{on} episodes are induced by the inward motion of clumps of cold gas into the centre of the galaxy, which provide a replenishment of new stars. The radiation drag or Poynting-Robertson effect caused by stellar radiation in the central star-forming region then dissipates the angular momentum of the gas and fuels the AGN. Our model suggests that reasonably powerful jet outbursts accompany each \qm{on} episode of the low-luminosity AGN, suggesting that, in analogy to the observed states of AGNs, during the active phases the Seyfert nucleus would be seen as radio-loud, becoming radio-quiet during the \qm{off} phase. In this section, we compare the properties of our simulated AGN with those of observed ones in disk galaxies.

%compare Lbol to observations (Ho+97)
%radio-loud episodes coincide with Seyfert episodes, Ho+01
In the reference model, the range of values obtained for the AGN bolometric luminosity for $z \lesssim 1$ agrees well with the median observational values for LINERs (median $L_{\rm bol} \approx 10^{41} \; {\rm erg}\ {\rm s}^{-1}$) and Seyferts (median $L_{\rm bol} \approx 10^{42} \; {\rm erg}\ {\rm s}^{-1}$) inferred from the Palomar spectroscopic survey of nearby galaxies \citep{ho04}. Since the Seyfert episodes of the central AGN coincide with jet outburst events during which the AGN would be seen as radio-loud in our simulations, this suggests that observed Seyfert/LINER nuclei would be mostly radio-loud objects rather than radio-quiet given their formation history, contrary to the generally held notion existent in the literature (e.g., \citealt{krolik99,laor00}). \citet{ho01} and \citet{tera03} however investigated several low-luminosity AGN and their radio-loudness classification, obtaining that a large fraction of Seyfert/LINER nuclei are radio-loud, a finding that corroborates our results. In particular, \citet{ho01} obtained that $\gtrsim 60 \%$ of the sources in their sample are radio-loud.

% kpc-scale effects of jets - compare with obs.
In our simulations, the numerical resolution is not sufficient to resolve the details of the propagation of the jet as it flows out of the AGN and how it transfer its power to the surrounding medium, for instance the spatial resolution is 0.5 kpc. As such, we assume that a fraction of the jet power is injected as thermal energy in the diffuse gas around the AGN on kpc-scales. The size of the region which is affected by the feedback from the AGN may be as large as $\approx 10-15$ kpc, although the jet itself is not resolved in the simulation. 
Two important questions arise: (a) are the heating effects due to jets generated in Seyfert galaxies able to affect kpc-scales and 
(b) can they be a source of feedback? Observations provides appealing clues to answering these questions.

% answer to question a
% outflow with small collimation: supported by surveys
% although our second model gives feedback energy preferentially along the disk axis, there is no reason for this: Schmitt+01
Several observational studies of Seyferts suggest that Seyfert jets may indeed reach kpc-scales and beyond. 
The Very Large Array surveys of \citet{colbert96} and \citet{galli06} find that extended kpc-scale radio outflows most probably associated with jets are common in Seyfert galaxies, although they cannot rule out entirely the possibility that these outflows arise from starburst-driven winds in some sources. These radio outflows have random orientations with respect to the host galaxy, a result which is maintained for jets observed from sub-kpc to kpc scales \citep{kinney00,schmitt01}, and have distorted, uncollimated morphologies. 
Therefore the observed radio outflows could provide a broad effective surface for depositing the jet power in the surrounding gas, as envisioned in our simulations. Several targeted observations of individual Seyfert galaxies also demonstrate that kpc-scale radio jets are not a rare feature in disk galaxies (e.g., \citealt{wilson83,morganti98,cecil00, whittle04,keel06,kharb06,middel07}). 

% answer to question b
% possible "bubble feedback" effect: Colbert+98, look for its citations (rupke+05 etc)
X-ray observations also reveal kpc-scale outflows of X-ray-emitting hot plasma in Seyfert galaxies \citep{colbert98,rupke05}, 
cospatial with the large scale radio outflows. 
Two favoured interpretations are that these kpc-scale X-ray outflows are generated either by the AGN jets that entrain and heat gas as they flow out of the nucleus, or through heating of the surrounding gas via thermalisation of the kinetic power of the jet, generating a wind. Either way, these processes bear resemblance to the generation of cavities or bubbles of X-ray emitting plasma in the central elliptical galaxies of clusters of galaxies, which are inflated by the radio jets (e.g., \citealt{dal04,bir04,fab06}). These observations therefore provide evidence for the possible importance of feedback due to the radio jets in Seyfert galaxies, which could heat the circumnuclear plasma and generate X-ray emitting outflows.

% comparison of jet powers
%   from radio luminosity (0.01): wilson, kharb, morganti (1e42-1e43)
%   also: capetti99, bicknell98 (mentioned by galli06)
Observational estimates of the kinetic power carried by the jets in Seyferts using different methods indicate that it is in the range $\sim 10^{42}-10^{43} {\rm erg}\ {\rm s}^{-1}$ (e.g., \citealt{wilson83,morganti98, bicknell98,capetti99,kharb06}). This observational range is in agreement with the values of the jet power which are generated during the radio-loud episodes for $z \lesssim 1$ in the simulation, $L_{\rm jet} \approx 10^{42}-10^{43} \; {\rm erg}\ {\rm s}^{-1}$ (Fig. \ref{RELATIVE}). 
It is worth noting that the jet power in our simulation is estimated using a physical model for the energetics of the jet, 
which depends on fundamental parameters of central engine, especially the accretion rate which is self-consistently calculated.

% efficiency of thermalization of jet power: open issue
% cite galli06 discussion
The fraction of the jet power generated by the AGN which is actually available to heat the gas on kpc-scales is a matter of debate. In the \sm{reference} simulation we adopt that a fraction $\eta_{\rm AGN} = 0.1$ of the jet power is deposited as thermal energy of the diffuse gas, although it is possible that this may be an overestimate of the value of $\eta_{\rm AGN}$. For instance, based on ram pressure and terminal velocity considerations of the lobes of kpc-scale radio outflows in Seyferts, \citet{galli06} argues that the Seyfert jets lose almost all their power within the inner kpc, implying that $\eta_{\rm AGN}$ may be considerably smaller than 0.1. 
The other simulations we present encompass the uncertainty in the value of $\eta_{\rm AGN}$. In this respect, the simulation adopting $\eta_{\rm AGN} = 0.01$ does not show any significant difference from the simulation without AGN feedback ($\eta_{\rm AGN} = 0$). 
On the other hand, when 100\% of the jet energy is transferred to the surrounding star forming gas (\sm{ism-fb} model), 
most of the feedback energy is rapidly radiated away, and the results are very similar to the \sm{reference} model 
in which 10\% of jet energy is given to the ambient halo gas. 
Thus, a better knowledge on the \qm{microscopic} physics of jet-ISM interactions is required in order to improve 
the understanding of the effects of AGN feedback in Seyfert galaxies.

%Stochastic cyclic activity (Sanders, Hopkins+06) 
Our simulations show that during the cosmological formation of a typical spiral galaxy, 
a low-luminosity AGN (Seyfert or LINER) is formed which has a stochastic cycle of activity, 
with many short-lived recurrent accretion events accompanied by jet outbursts. 
This finding agrees with the conclusion of \citet{sanders84} that the Seyfert activity is the result of short-lived stochastic 
accretion events. 
He arrived at this conclusion by considering plausible mechanisms that may fuel the central black holes of Seyferts, 
which are stochastic in nature, and our simulations track the evolution of the AGN by implementing one of these mechanisms: 
the accretion of molecular clouds to the galactic centre. Our findings therefore provide support to the assumption of 
\citet{hop06sy}, central to their work, that the central supermassive BHs of Seyferts stochastically increase their mass by 
accretion of cold gas.

%AGN lifetime (Woltjer, Sanders, Ho, Kharb)
%    If LINERs are included, lifetime increases to ~3 Gyr (Ho+97) 
%duration of episodes (Sanders, Kharb, Galli)
In the period $z \lesssim 1$ after the galactic disk is formed, the AGN is active only during a fraction $\approx 9\%$ 
of the its life in the \sm{reference} model, $\approx 25\%$ in the \sm{weak-fb} model and $\approx 1\%$ in the \sm{ism-fb} model.
This implies that the lifetime of Seyfert activity in the spiral galaxy is $\approx 10^9$ yr for the \sm{reference} or \sm{weak-fb} model and $\approx 10^8$ yr for the \sm{ism-fb} model. 
This is in agreement with statistical Seyfert lifetime $\sim 10^8-10^9$ yr estimated by counting the fraction of spiral 
galaxies that are observed to be Seyferts \citep{woltjer68,sanders84,ho97}. 
%The duration of each individual episode of activity in the simulation is typically $\lesssim 16.5$ Myr, 
%where this value corresponds to the time-step used to output the snapshots of the simulation. 
\citet{sanders84} estimated that each single episode of activity would not last longer than $10^6$ yr based on the size of Seyfert jets, 
a prediction which is in agreement with the detailed study of \citet{kharb06}. 
Unfortunately the time-step we used to output snapshot files is 16.8 Myr and therefore we may miss many of such short-lived events. 
Considering this, the \sm{ism-fb} model may be more comfortable than the \sm{reference} model because there is no long-lived accretion event in the \sm{ism-fb} model for $z \lesssim 1$ (lower-left panel of Fig.~\ref{DIFFFB}). 
It is also possible that the observed short duration of activity is not caused by the large-scale mass accretion to the galactic 
centres but by the physical processes in the accretion disk (e.g., instabilities) which we do not take into account in this paper. 

\section{Summary and Conclusions} 

In this paper, we have introduced a new methodology that enables us 
to simultaneously model star formation, SN feedback, BH accretion, 
and AGN feedback in cosmological simulations of galaxy formation. Our
treatment of black hole accretion is based on a theoretical model of the mass 
accretion into the galactic centre due to radiation drag. This connects the 
kpc-scale star formation activity in the galactic centre with the 
BH accretion rate.
Furthermore, motivated by the recent semi-analytic models
of \citet{cro06} and \citet{bow06}, we have distinguished two distinct accretion 
modes of BH fuelling, namely standard (optically thick, geometrically thin) 
Shakura-Sunyaev disks and (geometrically thick, optically thin) RIAFs.
This allows us to distinguish radio loud and radio quiet AGNs
depending on their mass accretion rate. 
In this model, we only consider AGN radio feedback in which we assume a fraction of the jet power 
is thermally coupled with the surrounding diffuse gas, and assume RIAFs are responsible for production of powerful jets.
As the first application of this model, we have performed cosmological simulations of the formation of a disk galaxy and 
its coevolution with the central AGN. 

Our results reveal a fundamental AGN-starburst connection during the evolution of the galaxy and the central AGN. 
During the starburst period ($z \gtrsim 1$) there is a lot of gas available to be accreted and the AGN is fuelled 
at high accretion rates, as a consequence the accretion disk is in the standard/thin state and the AGN is relatively 
luminous and produces weak jets (radio-quiet state). 
The starburst phase ceases at $z \sim 1$ because most of the gas at the galactic centre is
consumed by star formation or blown out by the associated SN feedback. Afterwards, the AGN fuelling rate drops and
the accretion disk then switches to a radiatively inefficient (RIAF) state. 
Due to a combination of small accretion rates and the low radiative efficiency of the RIAF, the AGN becomes 
a low-luminosity AGN or Seyfert nucleus which is almost invisible compared to the host galaxy. 
At this point even a little star formation activity around the central BH triggers intense jet production that regulates 
further gas cooling to the centre. Thus, in the quiescent star forming regime after $z \sim 1$ the AGN has a stochastic fuelling, 
undergoing several short-lived recurrent episodes of activity. During these accretion episodes the AGN brightness increases 
and jet outbursts are generated, which correspond to the radio-loud state. In-between these Seyfert episodes the nucleus 
is turned off and the galaxy is inactive.

Below $z \sim 1$ AGN feedback becomes important if 10\% of jet energy is available to heat the ambient diffuse 
halo gas (or 100\% is fed into the surrounding ISM).  
This powerful episodic AGN feedback suppresses star formation in the disc and almost completely halts the 
star formation in the bulge. 
This effect is minor when the wind speed due to SN feedback is fast compared to the circular velocity of the host halo 
because the SN feedback alone can suppress the star formation in the bulge reasonably well.  
When the wind speed is slow compared to the circular velocity of the host halo, the AGN feedback becomes more important.
  %Because of the higher star formation efficiency at high redshifts, 
This leads to 
the formation of an old, red elliptical galaxy in the \sm{slow-wind} model.

Many properties of the simulated low-luminosity AGN and its jets are in broad agreement with observations of Seyfert galaxies 
and their jets: namely its typical nuclear bolometric luminosity and radio luminosity, the coincidence of the radio-loud state 
with the Seyfert episodes, the fact that the jet may reach kpc-scales and strongly affect the surrounding gas on these scales, 
and the predicted AGN lifetime and duration of individual episodes of activity. 
The stochastic pattern of nuclear activity that emerges from our simulation seems to confirm the prediction of 
\citet{sanders84} for Seyfert galaxies.

The ratios between the central BH mass and the mass of the host spheroid at 
$z = 0$ in our simulations are $\sim 10^{-3}$ regardless of the strength of 
either SN or AGN feedback. 
This result is direct outcome of the radiation drag model which relates the star formation rate 
in the central star-forming region and the mass accretion rate to the central BH. 
The ratio is in good agreement with the observed ratio \citep{kor95, mag98, mer01, mcl02}, although 
the efficiency of the radiation drag we assumed to obtain this 
ratio is somewhat higher than the value suggested by \cite{ku02}. 
However, as we have discussed, the geometry and density structure of 
star forming regions can easily change the efficiency. 
We also note that recent semi-analytic models 
\citep{bau05, nag05a, nag05b} and cosmological simulations \citep{oka05}
suggest that stars may be born with top-heavy IMFs (and thus stronger 
radiation fields and higher radiation drag efficiencies) in starbursts.
The accretion rate by radiation drag can be much higher with this assumption
because stellar populations with top-heavy IMFs produce more radiation. 

The self-consistent treatment of BH accretion and associated
AGN feedback in galaxy formation enables us to explore 
the interplay between galaxies and AGNs in cosmological simulations 
of galaxy formation. 
The self-regulation of mass accretion due to feedback from RIAFs produces galaxies whose AGN 
properties are well matched to the observations.
The model is also qualitatively 
consistent with recent semi-analytic models that successfully reproduce 
many of the global
properties of the galaxy population. 
The effects of our AGN feedback implementation on formation of groups of galaxies, 
where more powerful AGN feedback from larger BHs is expected, will 
be presented in a forthcoming paper.

\section*{Acknowledgments}
We greatly appreciate the detailed reading and shrewd comments of the 
anonymous referee that strengthened this paper.
We thank Takayuki Saitoh, Nozomu Kawakatu, 
Yasuyuki Watabe, Carlos Frenk, Tom Theuns and Thaisa Storchi-Bergmann for useful discussions.
TO acknowledges financial support from the Japan Society for the 
Promotion of Science for Young Scientists (1089) under which part of 
this work was carried out.
RSN acknowledges support from the Brazilian institutions CNPq and CAPES, 
and by the European Commissions ALFA-II programme through its funding of 
the Latin-american European Network for Astrophysics and Cosmology (LENAC).
RGB thanks PPARC for the support of a senior research fellowship. 
We are grateful to Volker Springel for making the GADGET2 code public. 
All simulations were performed on the Cosmology Machine at the Institute 
for Computational Cosmology in Durham University. 
Part of analyses were carried out on GRAPE system at the Center for 
Computational Astrophysics, CfCA, of the National Astronomical 
Observatory of Japan (Project-ID: g06b13).

%\bibliography{okamoto}

\label{lastpage}
\end{document}